\pdfoutput=1
\documentclass[manuscript,screen,acmlarge]{acmart}
\usepackage{array}
\usepackage{tabularx}
\usepackage{amsmath}
\usepackage{hyperref}
\usepackage{booktabs}
\usepackage{longtable}
\usepackage{multicol}
\usepackage{multirow}
\usepackage[nameinlink]{cleveref}
\usepackage{makecell}
\usepackage{colortbl}  
\usepackage{xcolor}

\AtBeginDocument{%
  }

\setcopyright{acmlicensed}
\copyrightyear{2024}
\acmYear{2024}
\acmDOI{XXXXXXX.XXXXXXX}

\acmJournal{IMWUT}
\acmVolume{0}
\acmNumber{0}
\acmArticle{0}
\acmMonth{0}

\begin{document}

\title{Are We in The Zone? Exploring The Features and Method of Detecting  Simultaneous Flow Experiences Based on EEG Signals}

\author{Baiqiao Zhang}
\email{baiqiao@mail.sdu.edu.cn}
\affiliation{%
  \institution{Shandong University}
  \city{Weihai}
  \country{China}
}

\author{Xiangxian Li}
\affiliation{%
  \institution{Shandong University}
  \city{Jinan}
  \country{China}}
\email{larst@affiliation.org}

\author{Yunfan Zhou}
\affiliation{%
  \institution{Zhejiang University}
  \city{Hangzhou}
  \country{China}}
\email{zhouyunfan00@163.com}

\author{Juan Liu}
\affiliation{%
  \institution{Shandong University}
  \city{Weihai}
  \country{China}}
\email{zzzliujuan@sdu.edu.cn}

\author{Weiying Liu}
\affiliation{%
  \institution{Shandong University}
  \city{Weihai}
  \country{China}}
\email{2202237565@mail.sdu.edu.cn}

\author{Chao Zhou}
\affiliation{%
  \institution{Institute of Software Chinese Acadamy of Sciences}
  \city{Beijing}
  \country{China}}
\email{zhouchao@iscas.ac.cn}

\author{Yulong Bian}
\affiliation{%
  \institution{Shandong University}
  \city{Weihai}
  \country{China}}
\email{bianyulong@sdu.edu.cn}
\authornote{Yulong Bian is the corresponding author.}

\renewcommand{\shortauthors}{Zhang et al.}

\begin{abstract}
When executing interdependent personal tasks for the team’s purpose, simultaneous individual flow (simultaneous flow) is the antecedent condition of achieving shared team flow. Detecting simultaneous flow helps better understanding the status of team members, which is thus important for optimizing multi-user interaction systems. However, there is currently a lack exploration on objective features and methods for detecting simultaneous flow. Based on brain mechanism of flow in teamwork and previous studies on electroencephalogram (EEG)-based individual flow detection, this study aims to explore the significant EEG features related to simultaneous flow, as well as effective detection methods based on EEG signals. First, a two-player simultaneous flow task is designed, based on which we construct the first multi-EEG signals dataset of simultaneous flow. Then, we explore the potential EEG signal features that may be related to individual and simultaneous flow and validate their effectiveness in simultaneous flow detection with various machine learning models. The results show that 1) the inter-brain synchrony features are relevant to simultaneous flow due to enhancing the models' performance in detecting different types of simultaneous flow; 2) the features from the frontal lobe area seem to be given priority attention when detecting simultaneous flows; 3) Random Forests performed best in binary classification while Neural Network and Deep Neural Network3 performed best in ternary classification.



\end{abstract}



\begin{teaserfigure}
  \includegraphics[width=\textwidth]{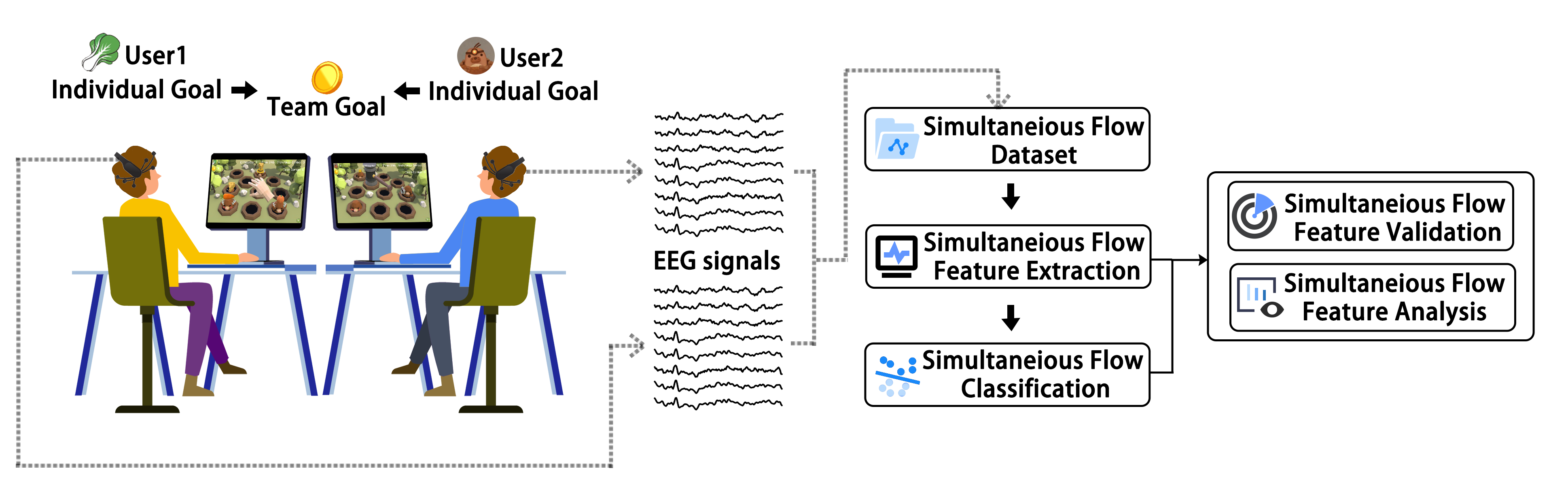}
  \caption{Detecting team members’ simultaneous flow experience during a two-user collaborative tasks based on multi-channel electroencephalogram(EEG) signals.}
  \label{fig:teaser}
\end{teaserfigure}


\maketitle
\section{Introduction}
\label{section1}
The flow experience (short as flow) is an important user experience during human-computer interaction\cite{10.1145/3173574.3173975}. It is a highly enjoyable mental state in which the individual is fully immersed and engaged in activities, and is thus considered the optimal experience\cite{2008Flow}\cite{csikszentmihalyi1990flow}\cite{sweetser2005gameflow}. Flow exists not only in performing individual activities, but also in performing collaborative activities. Team members may simultaneously experience flow and positive collective experiences while performing tasks for the team’s purpose, reflecting the unified and coordinated feelings generated among team members as they strive toward a common goal, which is known as team flow or group flow \cite{2016Group}. Team flow is considered to be the goal of optimal collaboration\cite{van2019TeamFlow}. Studies have shown that the experience of team flow is important for optimizing team performance: it can foster closer emotional connections between team members and the work environment, increase team experience and well-being at the team level, and thereby stimulate higher motivation, creativity, and both team and individual performance\cite{2016The}\cite{2012Flourishing}\cite{Bozanta2016serious}\cite{Keith2018Team}\cite{van2019TeamFlow}. As multi-user collaborative applications become more common, the team flow experience is increasingly focused on and considered an important user experience goal in collaboration and cooperation.

Although some researchers consider team flow as a solely group phenomenon \cite{2016Group}, van den Hout et al defined it as a concatenative experience of team members' individual flow \cite{van2018conceptualization}. Simultaneous individual flow while executing personal tasks for the team’s purpose may be the antecedent condition of achieving shared team flow\cite{boot2024we}. Considering shared team flow is a complex state and hard to operationalize, according to van den Hout et al.'s concatenative idea, we first focus on the antecedent condition and describe it as "simultaneous flow experience in collaborative task" (short as simultaneous flow), which means team members experiencing individual flow at the same time\cite{boot2024we}. Therefore, methods of detecting simultaneous flow are critical for better understanding team members' status, which is important for the dynamic optimization of multi-user collaborative systems.  
    
The current primary methods to evaluate flow are questionnaires, scales, interview and the experience sampling method (ESM)\cite{csikszentmihalyi2014experience}\cite{1996Development}\cite{peifer2021advances}. These methods are retrospective and have subjective response bias\cite{ye2020flow}. By contrast, physiological measurements have shown greater potential in evaluating flow in a more objective and real-time manner without disrupting the ongoing experience process\cite{bian2016framework}. For individual flow, recent studies have found correlations between flow and neural activity signals (e.g. frontal lobe activity)\cite{bruya2010effortless}\cite{nacke2010affective}, and explored flow detection methods based on electroencephalogram (EEG) signals\cite{chatterjee2016probabilistic}\cite{katahira2018eeg}\cite{plotnikov2012exploiting}\cite{wang2014exploratory}. These studies preliminarily demonstrate the effectiveness of this method. 
However, the methods of detecting simultaneous flow/team flow have not been investigated in greater detail. Therefore, this study focus on the simultaneous flow, exploring its related features and potential detect methods based on EEG signals.

We first considered the smallest team size and designed a two-user game to induce simultaneous flow. Based on the task, we recorded the paired participants’ multichannel EEG signals from brain regions related to flow experience. After labeling, slicing, and processing the data, we constructed the first multichannel EEG dataset with multiple labels of flow experiences in collaborative tasks. Next, we focused on inter-brain synchrony features from different brain regions' EEG signals and explored their effectiveness in detecting simultaneous flow by using several classifiers, including Logistic Regression (LR), Support Vector Machines (SVM), Decision Trees (DT), Random Forests (RF) and several Neural Network (NN) models. 

The main contributions of this research include:
\begin{itemize}
\item We designed a two-user task for continuously measuring simultaneous flow in collaboration based on which we constructed the first multichannel EEG dataset of simultaneous flow.
\item We proposed inter-brain synchrony features from EEG signals which potentially relevant to simultaneous flow for the first time. Based on comparison experiments, feature importance experiments and ablation study, we validated the effectiveness of inter-brain synchrony features in detecting simultaneous flow, and found features from frontal lobe area were given priority attention by the models.
\item We make the first attempt to recognize simultaneous flow based on EEG signals by using multiple machine learning methods. In this study, RF achieved the highest accuracy in binary classification (83.9\%); NN and DNN3 performed best in ternary classification (87.2\%). 


\end{itemize}

\section{Related Work}

\subsection{What are Team Flow and Simultaneous Flow?}
\label{section2.1}

Individual flow refers to a mental state characterized by perceiving oneself to be entirely absorbed in the current personal activity\cite{reducing2022Bian}\cite{csikszentmihalyi2000beyond}. Compared to individual flow, team flow (or group flow) places greater emphasis on the interactions among team members and achievement of an enjoyable experience in team level. Sawyer\cite{2016Group} was one of the first researchers to propose the concept of team flow. He argued that team flow depends on the interactions between team members and may manifest through such interactions. Compared with individual flow, three major differences were identified: (1) Team flow requires behavioral, cognitive, and emotional interactions among team members, whereas individual flow is generated from a solo activity\cite{2016The}; (2) In team flow, collective goals take precedence over other elements of the group, whereas individual flow is more focused on personal goals and outcomes; (3) Team flow requires a higher level of skill among team members\cite{2003Nuclear}. Shared team flow is a complex state of and more.

In more recent studies, while van den Hout et al. conceded that a group can attain a "collective state of mind", they disagree with the view that team flow allows for individuals not to experience flow\cite{van2018conceptualization}. Van den Hout et al. defined team flow as as a shared experience of flow derived from an optimized team dynamic during the execution of interdependent personal tasks. In this definition, "shared" means that individual team members are experiencing flow simultaneously and collectively while executing their personal tasks for the team's purpose(s)\cite{van2018conceptualization}. In other words, they view team flow as a concatenative experience, proposing that simultaneous individual flow while executing personal tasks for the team’s purpose may be the antecedent condition of achieving shared team flow. As we explained in the Section \labelcref{section1}, we introduce the term "simultaneous flow" to describe simultaneous flow experience in collaborative task, representing the state that team members experiencing individual flow at the same time \cite{boot2024we}. In this paper, we focus on "simultaneous flow" for the first step, and preliminary explore whether it has similarities with shared team flow.




\subsection{Flow Detection based on EEG Signals}

Because traditional flow evaluation methods (such as questionnaires and interviews) are retrospective and subjective, researchers have paid increasing attention to the potential of objective physiological signals in flow experience evaluations\cite{bian2016framework}\cite{cherep2022mental}\cite{ye2020flow}. Current studies on this topic primarily focus on evaluating individual flow experiences using physiological signals, such as electrocardiograms (ECG), electrodermal activity (EDA), and electroencephalograms (EEG). Based on these physiological signals with flow experience labels, supervised learning (through traditional machine learning or deep learning models) has been used to recognize and detect different flow experiences. Ye et al.\cite{ye2020flow} constructed several machine learning models to detect user gaming flow experiences using skin conductance and heart rate signals. They achieved an accuracy rate of 90\% in binary classification tasks by using a support vector machine classifier and achieved an accuracy rate of 61\% in ternary classification tasks by using random forest algorithm. In terms of deep learning, Maier et al.\cite{maier2019deepflow} proposed a DeepFlow model based on deep learning to classify high-level and low-level flow, as well as to classify different experiences of flow, boredom, and stress. Cherep et al.\cite{cherep2022mental} used wearable devices to collect EEG signals and used the deep learning model EEGNet to identify flow experiences with an accuracy of over 65\%. EEGNet uses raw multichannel EEG signals as inputs without pre-extracted features. According to research on the neural mechanisms of the team flow experience\cite{shehata2020team}, EEG signals contain valuable information regarding the flow experience. Existing research on flow metrics focuses primarily on the individual tasks, and lacks exploration of flow metrics at the collaborative tasks. Therefore, this study focuses on detecting flow experience in collaborative tasks based on EEG signals.

\subsection{Physiological Synchrony}

Chatterjee et al. pointed out that the phenomenon in which the physiological responses of two individuals become more similar is called "physiological synchrony"\cite{chatterjee2023automated}. Research on the neural activities of team flow showed that team flow is characterized by higher inter-brain synchrony in the left temporal cortex\cite{shehata2020team}, revealing the neural activity mechanisms of team flow in collaborative tasks based on neural physiological signals. 
Current research on physiological synchrony in detecting team flow/simultaneous flow has not yet been conducted. Therefore, we begin by referring to relevant research in affective computing and summarize the following three important aspects.

\subsubsection{Experimental Tasks for Physiological Synchrony}
In terms of experimental design, studies on physiological synchrony commonly require the group participants to engage in collaborative tasks. Kevin et al. asked participants to play different roles as a team in a flight operation task and collected their EEG signals to detect the cognitive load and cooperation level among the team members\cite{verdiere2019spectral}. Darzi et al. measured each participant’s physiological response, physiological connectivity indices, and task performance under different conditions\cite{darzi2021automated}. They suggested that in multi-user settings, physiological synchrony information related to the specificity of interactions between participants can be obtained by examining the similarities in their physiological responses. The degree of linkage increases with the number of collaborations, the intensity of competition, and the sheer amount of joint attention. These findings can serve as a reference for designing simultaneous/team flow tasks.

\subsubsection{Features of Physiological Synchrony}
In terms of feature selection, the exploration of inter-brain synchrony features mainly focused on neural response consistency across participants. Chatterjee et al. categorized features into individual features, computed from the physiological data of a single participant, and synchrony features, computed from the physiological data of two participants\cite{chatterjee2023automated}. Darzi et al. constructed a feature set by combining individual physiological features with physiological connectivity features\cite{darzi2021automated}. Javier et al.\cite{Hernandez_Riobo_Rozga_Abowd_Picard_2014} found that a combination of features extracted from the child's EDA activity and physiological synchrony between the child and the adult resulted in the highest classification accuracy. Shkurta et al.\cite{Gashi_Di} computed the synchrony features from EDA signals using dynamic time warping (DTW) algorithm. The results show that an increment in physiological synchrony among students during a lecture is related to an improvement in students' emotional state. Studies using multi-person EEG data show that inter-brain correlations in theta and alpha wave amplitudes in the right temporoparietal junction and alpha and beta wave amplitudes in the frontal region are associated with understanding others’ intentions and high-level cooperative strategies\cite{shehata2020team}. Ding et al.\cite{ding2018inter} argued that inter-brain amplitude features and a combination of all inter-brain features outperformed single-brain features in the most competent participant. Inter-brain synchrony features should be considered in this study.

\subsubsection{Recognition Performance Based on Physiological Synchrony Features}
Ding et al.\cite{ding2018inter} used a method that combines physiological synchrony indices and classification algorithms. Using machine learning techniques such as linear discriminant analysis and random forests, they performed a four-category classification task and ultimately achieved a classification accuracy of 75\%. When the synchrony features were removed, the accuracy decreased to 65.6\%. Darzi et al.\cite{darzi2021automated} used the physiological responses from paired participants to automatically classify emotions within a competitive context. They used linear kernel support vector machines and ensemble decision trees to classify the participants’ psychological states using multiple physiological signals such as skin conductance, achieving a binary classification accuracy of 84.3\% and a ternary classification accuracy of 60.5\%. Kevin et al.\cite{verdiere2019spectral} assigned different roles and tasks to participants within the same scenario and constructed a linear discriminant classifier based on spectral EEG features to categorize different psychological states, achieving a classification accuracy of 60\% for team cooperation levels. Therefore, inter-brain synchrony features should be included in simultaneous flow detection.

In summary, neural-activity signals reflect flow experiences and can be effectively used to objectively evaluate them. This research focuses on EEG signals, fully considering the physiological synchrony features of team members, and explores simultaneous flow detection methods based on EEG signals. As there is currently no specialized, labeled EEG dataset for simultaneous flow, we begin by constructing an EEG dataset for simultaneous flow.

\section{Construction of Simultaneous Flow EEG Dataset}

In this study, we start with the smallest team size (two team members) to construct the first EEG dataset for simultaneous flow. Before collecting data, we need to first design a two-user collaborative experimental task to induce simultaneous flow.

\subsection{Design of An Experimental task for Inducing Simultaneous Flow}

\subsubsection{Design Rationale}
Currently, no unified task paradigm is specifically designed to induce simultaneous flow. To construct a simultaneous flow task, we first summarized the commonly used tasks in inducing individual flow, including "Whack-a-Mole," "Tetris," and other game tasks\cite{berta2013electroencephalogram}\cite{bian2016framework}\cite{harmat2015physiological}\cite{ye2020flow}, as well as mental arithmetic, chess, and learning quizzes, among others\cite{alzoubi2012detecting}\cite{katahira2018eeg}\cite{wang2014exploratory}. After a comprehensive comparison of these flow-inducing tasks, we designed the "Two-player Collaborative Whack-a-Mole Game" based on the traditional single-player Whack-a-Mole task. The main selection criteria were as follows:

\begin{itemize}
\item Game rules and operations are simple and provide high accessibility to various user groups.
\item The game can easily be designed as a collaborative task involving interdependent personal tasks.
\item The game can ensure high fluency in an operating task because it is less susceptible to random operational errors (e.g., unlike Tetris, in which a small error can accumulate a large effect).
\item The game requires the participants to be fast and highly focused, facilitating the rapid induction of the flow experience.
\item The difficulty of task can be dynamically adjusted in an easy way.

\end{itemize}

Finally, we developed an experimental system using the widely used game engine Unity 3D based on the design described in subsequent sections. 

\subsubsection{Storyline}
Previous studies demonstrated that the use of themes and storylines can enhance users’ involvement and intentional engagement\cite{berson2018excursion}\cite{bouchard2019virtual}. The storyline for this game is as follows:

You and your partner run a farm as a team by producing and selling agricultural products such as vegetables. A group of moles is trying to steal agricultural products. You and your teammates need to work together to protect your agricultural products and fight the moles to minimize losses and maximize profits.

\subsubsection{Task Goal and Collaboration Pattern}
According to van den Hout et al's conceptualization of flow\cite{van2018conceptualization}\cite{van2019TeamFlow}, the task Goal and collaboration Pattern were designed as follows.

\textit{Task Goal:} When team members aim to achieve a common team goal and pursue optimal team performance, they cooperate and collectively focus on their own personal task, thereby inducing simultaneous flow or event team flow. The goal of the team task is to collaboratively protect as many vegetables and eliminate as many moles as possible in a limited amount of time to maximize profits.

\begin{figure}[h]
  \centering
  \includegraphics[width=\linewidth]{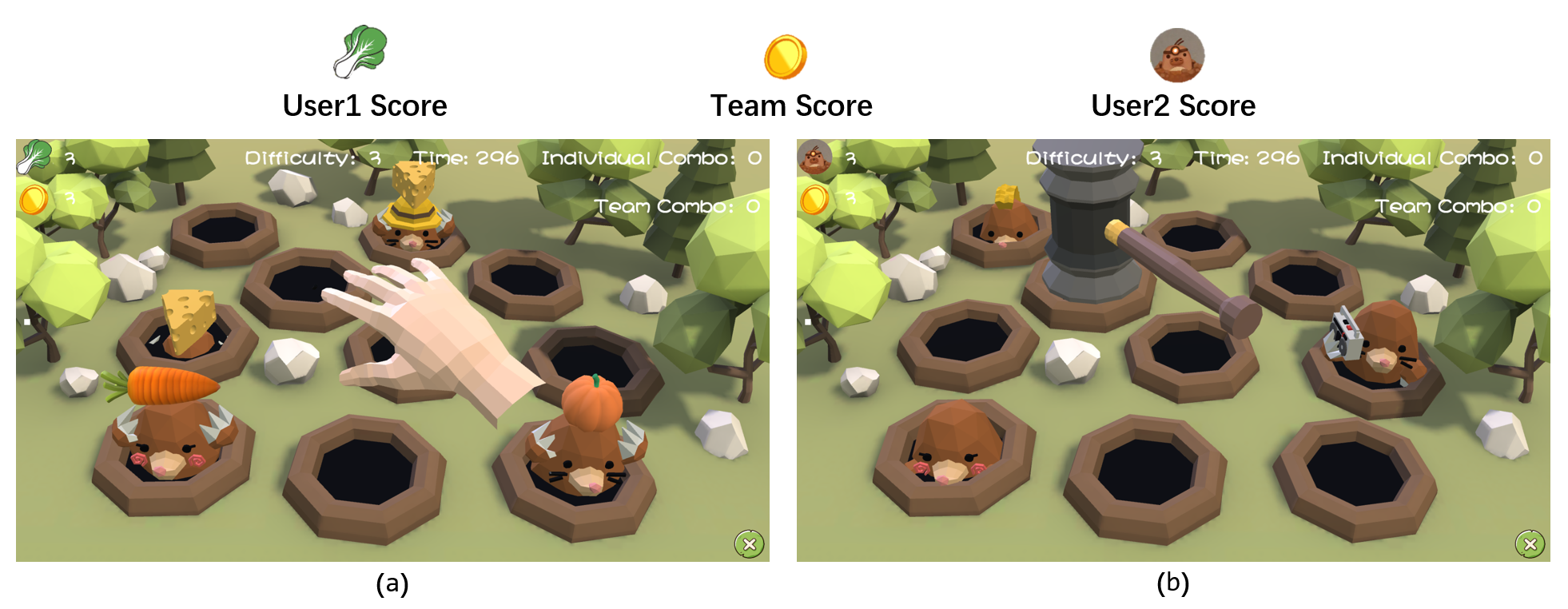}
  \caption{Screenshots of the collaborative game interface for user 1 (a) and user 2 (b). The interface displays various parameters related to team tasks, individual performance, and team performance in real-time.}
  \label{fig:mole}
\end{figure}

\textit{Collaboration Pattern:}
User 1 is responsible for hitting as many moles as possible to recapture the vegetables (Fig. \ref{fig:mole}a), whereas user 2 is responsible for hitting as many moles as possible to catch them (Fig. \ref{fig:mole}b). Although the division of labor in the games is different, the operational methods are identical. Individual performances are displayed on a user interface in real time.

Hitting targets continuously as much as possible can result in additional rewards for team profits. Team performance and profits are also displayed on the user interface in real time (Fig. \ref{fig:mole}). Each user’s performance in the personal task contributes to the team’s performance, which makes them collaborate as a team.

\subsubsection{Flow Experience Sampling}\label{subsubsec:3.1.4}
Different from previous studies which measured flow only once after the task, this study aimed to capture more immediate instances of flow. Therefore, we employed the Experience Sampling Method (ESM) to sample flow experience multiple times during the task, which is a common method in flow research\cite{peifer2021advances}. Each round of the game lasts for 5 min. After the game starts, the user experience is sampled every 1 minute (five times in total). The system was designed to pop out an evaluation interface by asking participants to score their individual flow experiences. Moreover, the game screen remains visible when the evaluation interface is displayed, which helps minimize disruption to the continuity of game experience.

\begin{figure}[h]
  \centering
  \includegraphics[width=\linewidth]{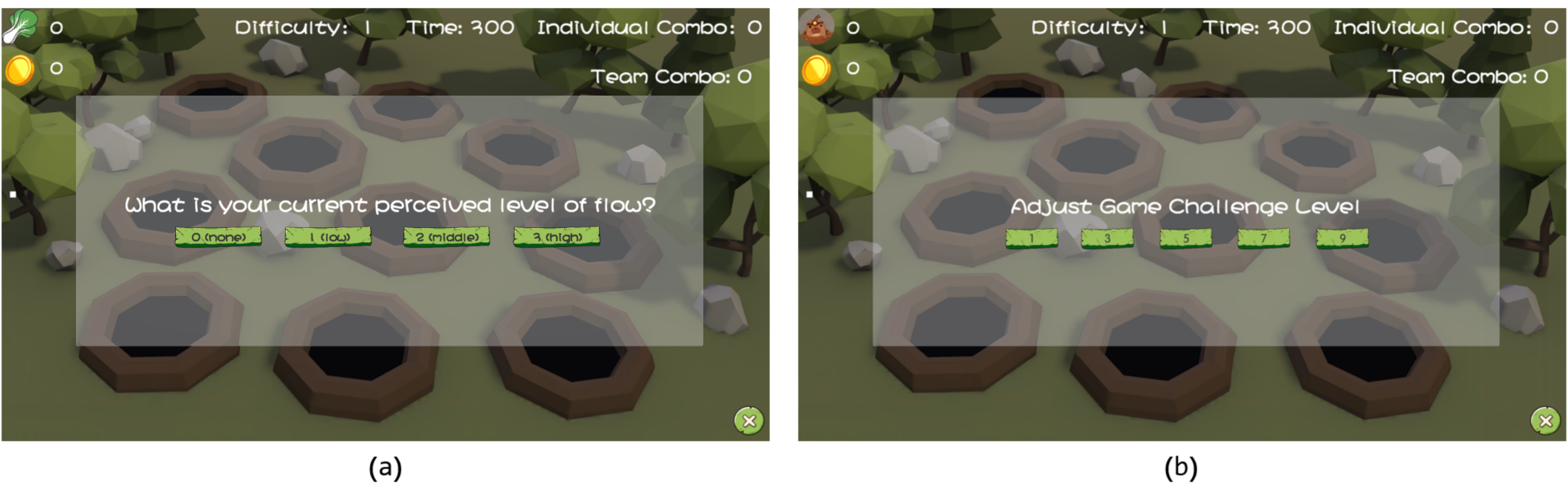}
  \caption{During the task, flow experiences are sampled at intervals by popping out an evaluation interface (a). Following the sampling, the game difficulty can be adjusted according to the two player’s skill by popping out an adjustment interface (b).}
  \label{fig:adjust}
\end{figure}

As to the measure of flow, when implementing ESM, repeatedly answering lengthy flow questionnaires will interrupt the flow state during the playing session and decrease the validity of the measurements\cite{perttula2017flow}, thus single-item measure is an effective way to minimize disruption to the continuity of flow experience\cite{kiili2021flow}. The effectiveness of the single-item measure has been supported by previous studies\cite{rodriguez2018development}\cite{ronimus2014children}. In this study, we adopted a single-item scale to evaluate the flow experience on a Likert scale ranging from 0 (no flow) to 3 (high degree of flow)(see Fig. \ref{fig:adjust}). The item was developed from a scientific questionnaire by Csikszentmihalyi to measure flow experiences \cite{peifer2021advances}\cite{rotgans2018individual}\cite{perttula2017flow}\cite{kiili2021flow}. According to the established procedure for using this questionnaire, we provided participants with a one-paragraph description of the flow concept and examples to ensure their comprehensive understanding before entering the game.

To enhance the rationality and effectiveness of the single-item measure in ESM, incorporating a test-retest procedure is beneficial to minimize the potential measurement error of using single-item measures\cite{dejonckheere2022assessing}. Accordingly, this study included a procedure where users reviewed the video recording immediately after completing the game to verify the accuracy of the scoring.

\subsubsection{Adjustment of Game Difficulty}
According to the classic flow-channel model, achieving flow experience requires a match between the individual’s skill and the level of task challenge (optimal challenge–skill balance)\cite{csikszentmihalyi1990flow}\cite{ye2020flow}. Appropriate difficulty is a critical factor for inducing a flow experience. Therefore, a mechanism for adjusting the difficulty of flow-inducing game tasks is necessary. 

To our knowledge, there is currently no objective method that can accurately and adaptively adjust the difficulty for paired participants. Therefore, considering the existing methods,  quickly communicating and reaching an agreement between two players could be a reliable approach. In this game, the optimal task difficulty is achieved by allowing the two users to actively adjust the speeds of the moles (time of each occurrence). There are four moles per occurrence and the speed is set at five levels. During each user experience sampling, the participants can quickly communicate whether to adjust the difficulty level, as shown in Fig. \ref{fig:adjust}(b).

\subsubsection{Recording of Game Data}
The system continuously records data for both users throughout the game, including: (1) the time points of game start, experience sampling, and game over; (2) response times for temporal alignment and segmentation with EEG signals; (3) rating scores of flow experience from sampling; and (4) game performance data.

\subsection{Acquisition of EEG Data }

\subsubsection{Participants}
52 pairs of participants were recruited from a local university to participate this study (mean age: 18.35 ± 1.81 years). None of the participants experienced alcohol consumption, excessive fatigue, medication use, or illness from the day before the experiment until the day of testing. Each pair of participants was randomly assigned to one of two roles: recapturing vegetables or catching moles. A total of 97.6\% of the participants were right-handed, 96.97\% had previously played video games, and 84.69\% were familiar with their teammates. The study was conducted in accordance with the guidelines of the Declaration of Helsinki and approved by the Human Research Ethics Committee of the local hospital. Informed consent was obtained from each participant.

Note: In this paper, we used different words to describe participants, such as "user", "player" and "team member". Their meanings are the same and we only use different expressions in different contexts.

\subsubsection{Data Acquisition Apparatus and Environment}
\begin{figure}[h]
  \centering
  \includegraphics[width=0.75\linewidth]{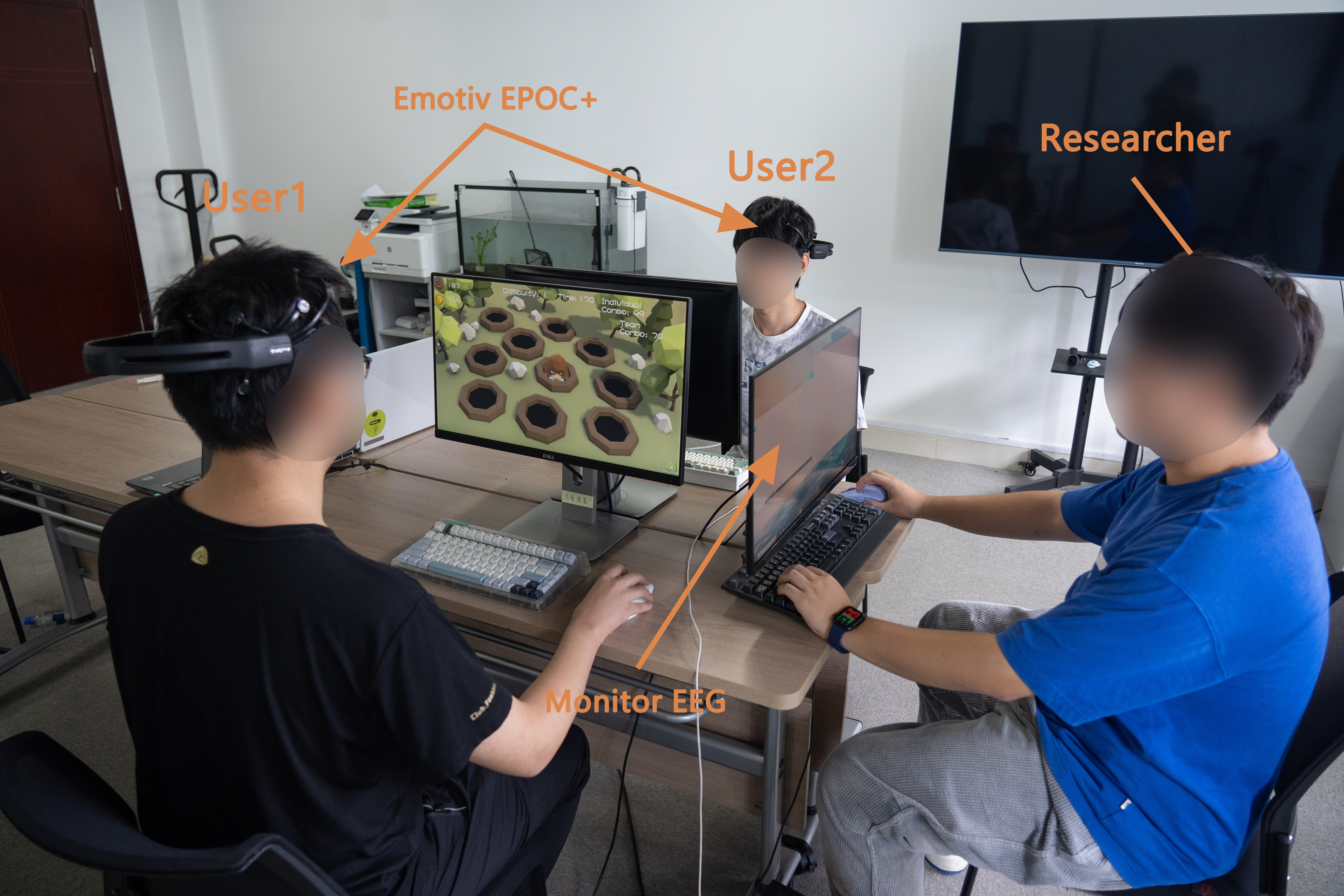}
  \caption{Data acquisition apparatuses and experimental environment.}
  \label{fig:experiment}
\end{figure}

To acquire EEG data of simultaneous flow, each pair of participants was required to perform the collaborative task simultaneously. They used two PCs (Legion Tower 7i Gen 7 with RTX 3070 and Dell monitors) with the same model to play the game (Fig. \ref{fig:experiment}), and performed the tasks using mice of the same model (Razer DeathAdder V3 Pro).

\begin{figure}[h]
  \centering
  \includegraphics[width=0.80\linewidth]{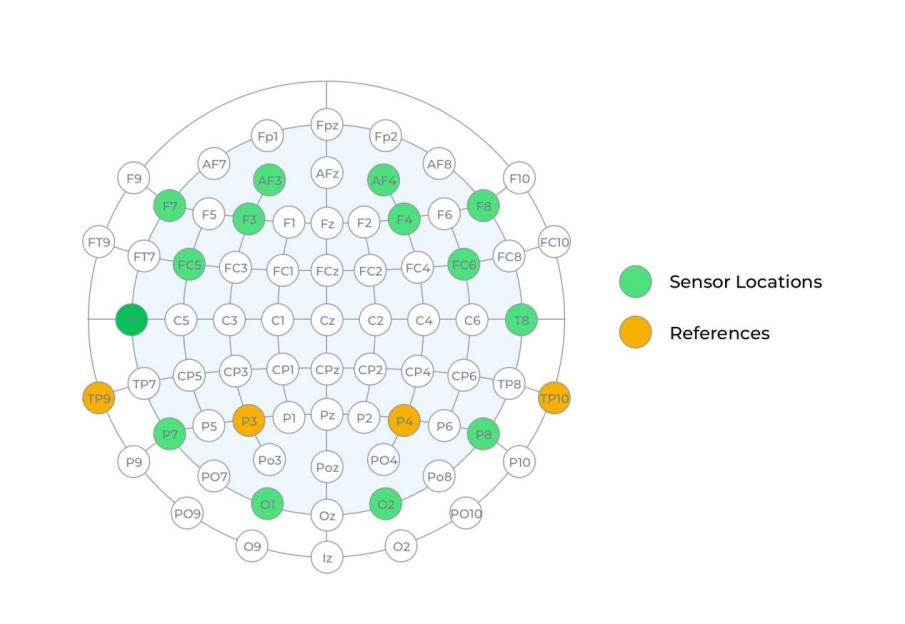}
  \caption{Electrode placement distribution for EEG data acquisition.}
  \label{fig:placement}
\end{figure}

Each participant wore an EEG recording device to acquire the EEG signals throughout the collaborative task. The EEG recording devices adopted in this study were two sets of Emotiv Epoc+, which supports the acquisition of up to 14 channels of EEG signals at a sampling rate of 256 Hz\cite{Liu2018Real-Time}. The software used for data acquisition is Emotive PRO. The electrode distribution for the 14-channel EEG signals follows the 10-20 international system, as shown in Fig. \ref{fig:placement}. We chose the Emotiv Epoc+ because it is easy to use with increased practicability and less physical restriction. Moreover, it has been effectively used in building EEG datasets for flow experience, emotions, and psychological behaviors\cite{McMahan2015Evaluating}\cite{Zhang2023SparseDGCNN}\cite{Katsigiannis_Ramzan_2018}\cite{Pham_Tran_2012}\cite{Taylor_Schmidt_2012}.

\textbf{}

\subsubsection{EEG Data Acquisition Procedure}
\begin{figure}[h]
  \centering
  \includegraphics[width=0.80\linewidth]{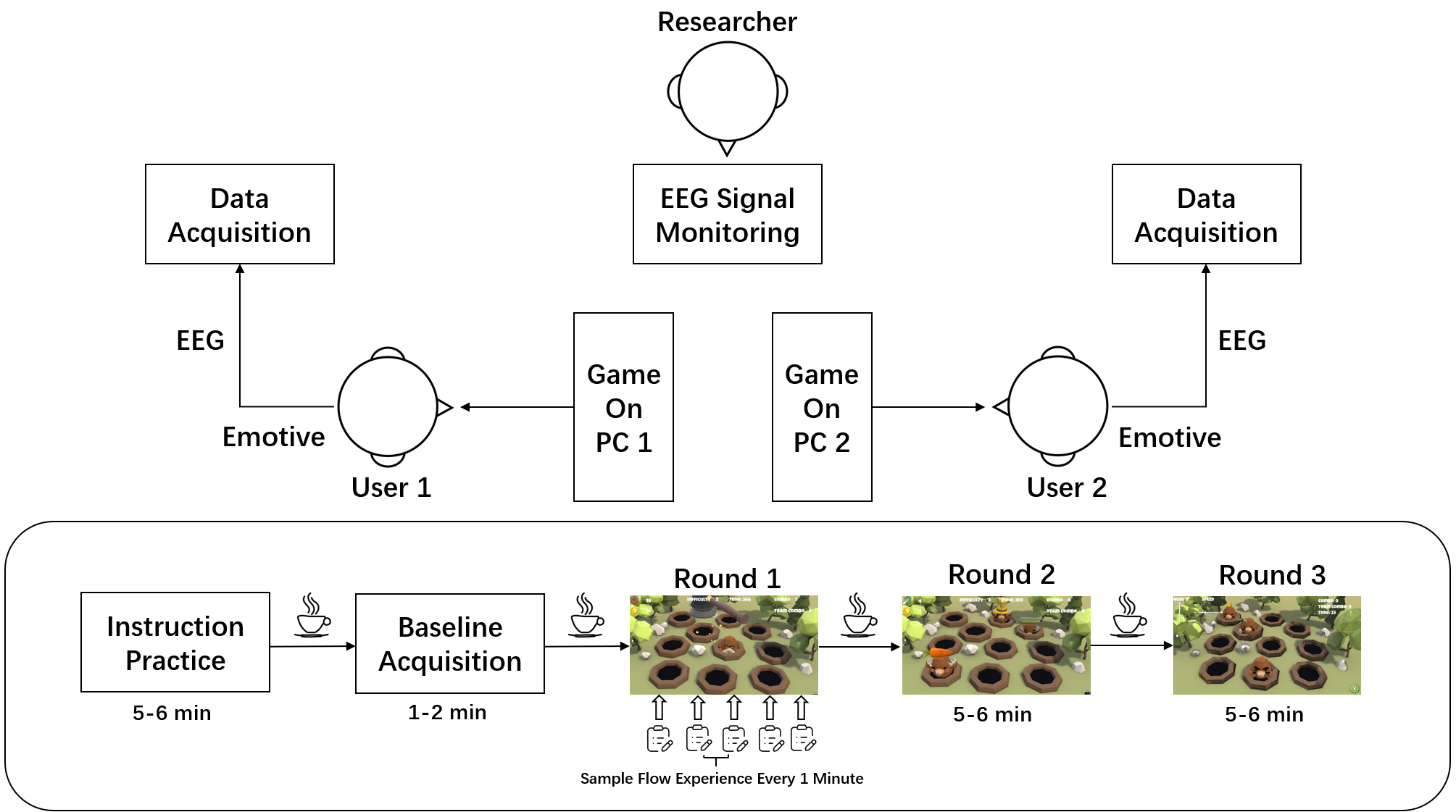}
  \caption{Process Diagram of Simultaneous Flow Collection System.}
  \label{fig:acquisition}
\end{figure}

The data acquisition process is shown in Fig. \ref{fig:acquisition}. 

First, the researcher explained the game details to each participant pair. Subsequently, each pair of participants practiced the operations in a practice scene. They then wore the EEG acquisition devices and entered a relaxed state. During this period, their baseline EEG data were collected for approximately 1 min. The participants then played three rounds of the game, with sufficient rest between rounds. To capture various flow-related experiences, the starting difficulty levels for the three rounds were varied. Flow experience sampling (introduced in Section \ref{subsubsec:3.1.4}) was performed during this period. During the game, the participants were asked to avoid head movements and facial expressions to reduce artifacts in the EEG signals. The researcher monitored the participants’ EEG data on a separate display throughout the game. If there were signal abnormalities during the process, they were recorded and considered to exclude the corresponding data from subsequent processing. It took approximately 30–40 minutes for each pair of participants to complete the EEG data acquisition process. Each participant obtained a corresponding course credit as a reward. Finally, we labeled and prepared the recorded EEG sample data for further dataset organization based on the flow scores obtained during user experience sampling.

\subsection{Dataset Organization}
\subsubsection{Selection of EEG Channels}
It has been demonstrated that the flow experience is associated with the activity in the frontal lobe of the brain\cite{katahira2018eeg}\cite{peifer2021advances}. Moreover, Shehata et al.\cite{shehata2020team} found that the left temporal lobe acts downstream in causal information relationships, receiving and integrating information from brain regions related to individual flow and social interaction, and participating in higher-order inter-brain neural synchrony\cite{csikszentmihalyi2014experience}. This area is activated during the emergence of team flow and plays a role in generating shared flow experiences\cite{shehata2020team}. Therefore, we have selected the EEG signals from he frontal and left temporal lobes to construct the simultaneous flow dataset. Finally, according to the distribution of the electrodes shown in Fig. \ref{fig:placement}, eight-channel EEG signals were selected: F3, F4, F7, F8, AF3, AF4, T7, and P7.

\subsubsection{EEG Signal Slicing}
For each participant, continuous EEG signals were recorded during each round of the game for 5 min. The data from the 6-second intervals preceding each sampling point were used for EEG data slicing. The form of data stored in each sliced signal file was channels × data points. The sampling rate is 256Hz, the number of data points in each file is 1536. We selected 6-second segments of EEG signals, due to a 6-second time window size strikes a balance between accuracy and computational efficiency. It is long enough to capture sufficient information of a subjective state\cite{Zhang2023SparseDGCNN}, and while being short enough to reduce computational complexity for dynamic computing\cite{candra2015investigation}. 


\subsubsection{Dataset Composition}
Although 52 pairs of participants participated in EEG data acquisition, 47 pairs were included in the final dataset after excluding groups with data acquisition errors and operational failures. The size of the dataset was approximately 34.4 GB, and contained EEG signals recorded 6 s before each flow experience sampling. All data were saved as CSV files, each of which was a matrix of dimensions [14, 1536]. These data represent 14-channel EEG data with 1536 sample points; the sampling rate of the EEG signals is 256 Hz i.e., each CSV file has a duration of 6 seconds.

During the game, the EEG signals collected from each pair of participants were organized by "group number- participant number-game set number.” Each participant’s folder contained EEG data sliced from five game rounds. 

The data and code of this paper is released at : \href{https://anonymous.4open.science/r/Team-Flow-Dataset-494E}{https://anonymous.4open.science/r/Team-Flow-Dataset-494E} anonymously. Detailed information about the data is also summarized in the "README.md" file.

\section{Feature Extraction for Individual and Simultaneous Flow} 
According to previous studies, EEG signals contain both individual\cite{chatterjee2016probabilistic}\cite{katahira2018eeg}\cite{wang2014exploratory} and simultaneous flow features\cite{shehata2020team}. Based on previous studies, we have summarized the EEG signal features related to individual flow and proposed features that may reflect inter-brain synchrony among team members for simultaneous flow. Methods for data denoising and feature extraction are introduced in this section.

\subsection{Data Denoising}
Raw EEG data contain noise and redundancy, thus requiring initial preprocessing. The main sources of noise in EEG signals include physiological artifacts, power line interference, and interference generated by activities such as blinking and facial muscle movements. Considering the frequency range of EEG signals and the frequency range of noise interference, this study employed wavelet threshold denoising methods to denoise each channel of the raw EEG signals.

\subsection{Feature Extraction and Normalization}
\label{featureExtraction}
Based on previous research analyzing the EEG signals of individual flow experiences, this study extracted some time- and frequency-domain features that may be relevant to individual flow experiences. Moreover, we preliminarily extracted EEG features related to inter-brain synchrony as potential simultaneous flow features. Detailed explanations are as follows.

\subsubsection{Features for Individual Flow}
\label{featurename}
This study extracts time and frequency-domain features that may be related to individual flow experiences, as listed in Appendix Table \labelcref{tab:features}. The time-domain features include the Mean, Standard Deviation(SD), Variance, and Average Absolute First-Order Difference(AAFOD), Normalized First-Order Difference(NFOD), Energy, Power, Hjorth Parameters(Activity, Mobility), Higher-Order Zero-Crossing(HOZC), Peak-to-Peak Mean(PPM), and Kurtosis.

Frequency-domain features mainly involve wavelet decomposition\cite{mallat1989theory} of the collected EEG signals into four non-overlapping sub-bands: \(\delta\) (0--4Hz), \(\theta\) (4--8Hz), \(\alpha\) (8--16Hz), and \(\beta\) (16--32Hz). 

The frequency-domain features are then extracted from these sub-bands in the eight EEG channels. Power Spectral Density(PSD) and  Logarithmic band power(LBP) are calculated for each of the four bands. Differential Entropy(DE) features are extracted from four frequency bands(\(\delta\), \(\theta\), \(\alpha\), and \(\beta\)) and the full frequency band (FB) of EEG signals in each channel. This study employs the Welch method\cite{welch1967use} to extract the power spectral density of four frequency bands: delta, theta, alpha, and beta, and computes the average of the power spectrum across channel dimensions.

\subsubsection{Features for Simultaneous Flow}
In the context of simultaneous flow, this study extracted inter-brain synchrony features from the EEG signals of each pair of participants, including cross-correlation coefficients and Dynamic Time Warping (DTW) distance.

Let \( A = \{a_1, a_2, a_3, \ldots, a_m\} \) represent the EEG signal values sampled from Participant 1 at a certain electrode, and \( B = \{b_1, b_2, b_3, \ldots, b_n\} \) represent the EEG signal values sampled from Participant 2 at the same electrode. The features are introduced as follows:

\begin{enumerate}
\item \textit{Cross-correlation Coefficient:}
The Pearson correlation coefficient for the brainwave signals between each channel and the frequency band of the two participants are computed\cite{hernandez2014using}. This serves as an index for assessing the signal correlation between team members.

\begin{equation}
\rho_{A, B} = \frac{\text{cov}(A, B)}{\sigma_A \sigma_B}
\end{equation}

Where \( \text{cov}(A, B) \) calculates the covariance between the two signals, \( \sigma_A \) is the standard deviation of EEG signal A, and \( \sigma_B \) is the standard deviation of EEG signal B.

\item \textit{Dynamic Time Warping Distance\cite{chatterjee2023automated}\cite{darzi2021automated}\cite{muszynski2018aesthetic}:}
This is a technique based on dynamic programming for quantifying the similarity between two signals, and is robust to time decorrelation. The specific computational steps are as follows:
\begin{enumerate}
    \item Calculate the Euclidean distance \( D(a_i, b_j) \) between every two sample points in the two signal sequences, where \( 1 \leq i \leq m \), \( 1 \leq j \leq n \).
    \item Find the shortest path distance from \( (a_1, b_1) \) to \( (a_m, b_n) \). The next node from a current node \( (a_i, b_j) \) must be among \( (a_{i+1}, b_j) \), \( (a_i, b_{j+1}) \), or \( (a_{i+1}, b_{j+1}) \), and the path must be the shortest.
    \item During each iteration to reach \( (a_i, b_j) \), follow the dynamic programming paradigm to choose the shortest distance to \( (a_i, b_j) \) from among \( (a_{i-1}, b_j) \), \( (a_i, b_{j-1}) \), or \( (a_{i-1}, b_{j-1}) \).
\end{enumerate}
\begin{equation}
D(a_i, b_j) = \text{Dist}(a_i, b_j) + \min \left\{ D(a_{i-1}, b_j), D(a_i, b_{j-1}), D(a_{i-1}, b_{j-1}) \right\}
\end{equation}

\end{enumerate}

\subsubsection{Feature Normalization}

Due to the typically significant variability in the feature values of EEG signals among different participants, this study employs the Z-score normalization method to standardize the features of all participants. This strategy aims to mitigate the impact of inter-individual differences in the range and variability of physiological signals\cite{ye2020flow}. Let $f$ represent the feature values extracted for all participants on a given feature, $\bar{f}$be the mean value of these feature values, and $f_{std}$ be the standard deviation of the feature values. The normalized value of the feature is then given by:
$$z = \frac{f - \bar{f}}{f_{std}}$$

\subsubsection{Overview of Features}
The individual and inter-brain synchrony flow features selected in this study are listed in Appendix Table \labelcref{tab:features}. By extracting features from the eight-channel EEG signals, 272 features (208 features of individual flow and 64 features of inter-brain synchrony) were selected, constituting the set of individual and simultaneous flow features used in this study.

\section{Validation of Flow Features}
\subsection{Purpose}
The primary purpose of this section is to validate the effectiveness of individual and inter-brain synchrony features in detecting different types of simultaneous flow and to analyze the impact of these features in this process. We began with validating the effectiveness of these features in binary and ternary classification tasks.

\subsection{Dataset Organization}

Based on the flow rating scores on EEG dataset, we established two sets of classification labels. The first set distinguishes between "Low Simultaneous Flow" and "High Simultaneous Flow." The second set categorizes the data into three groups: "Simultaneous Flow", "Individual but not Simultaneous Flow," and "Neither Individual nor Simultaneous Flow".
\subsubsection{Binary Classification Labels}
The binary classification labels of "Low Simultaneous Flow" and "High Simultaneous Flow" are co-determined by the flow experiences of the paired users. According to the definition of simultaneous flow \cite{boot2024we}, this study operationally defined it as "simultaneous flow occurs only when both participants achieve a high flow experience in collaborative task". Scores of "0" and "1" represent low flow experiences, while scores of "2" and "3" represent high flow experiences. Taking the flow experience of participant 1 as an example, the labeling rules of the Simultaneous Flow Labels are listed in Table \labelcref{table2}, and the distribution of sample labels in the data set is listed in Table \labelcref{table3}.

\begin{table}[h]
    \centering
    \caption{Simultaneous Flow Labels based on the two users’ flow score (Binary Classification)}
    \label{tab:flow_comparison}
    \begin{tabular}{c c c}
        \toprule
       Score of participant 1 & Score of participant 2  & Binary Classification labels \\
        \midrule
        0 & 0/1 & Low Simultaneous Flow \\
        1 & 0/1 & Low Simultaneous Flow \\
        2 & 2/3 & High Simultaneous Flow \\
        3 & 2/3 & High Simultaneous Flow \\
        \bottomrule
    \end{tabular}
    \label{table2}
\end{table}

\begin{table}[h]
    \centering
    \caption{Sample Label Distribution (Binary Classification)}
    \label{tab:sample_label_distribution}
    \begin{tabular}{c c c}
        \toprule
        Simultaneous Flow Label & Sample Count & Proportion (\%) \\
        \midrule
        Low Simultaneous Flow  & 512         & 37.1           \\
        High Simultaneous Flow & 868         & 62.9           \\
        \bottomrule
    \end{tabular}
    \label{table3}
\end{table}

\subsubsection{Ternary Classification Labels}
According to the definition of simultaneous flow, the ternary classification labels are co-determined by the paired participants' flow experience. Taking the flow experience of Participant 1 as an example, the labeling rules are listed in Table \labelcref{table4}. The distribution of the ternary classification sample labels in the dataset is presented in Table \labelcref{table5}, revealing the problem of sample imbalance.

\begin{table}[h]
\centering
\caption{Comparison of Individual Flow Ratings and Simultaneous Flow Labels (Ternary Classification)}
\label{tab:individual_team_flow_comparison_ternary}
\begin{tabularx}{\linewidth}{ m{60pt}<{\centering} m{60pt}<{\centering} m{100pt}<{\centering} X<{\centering}}
        \toprule
        Individual Flow of Participant 1 & Individual Flow of Participant 2  & Simultaneous Flow       & Ternary Classification Labels of Participant 1  \\
        \midrule
        Low  & Low & Low Simultaneous Flow     & Neither Individual nor Simultaneous Flow \\
        Low  & High & Low Simultaneous Flow    & Neither Individual nor Simultaneous Flow \\
        High & Low & Low Simultaneous Flow     & Individual but not Simultaneous Flow     \\
        High & High & High Simultaneous Flow   & Simultaneous Flow    \\
        \bottomrule
\end{tabularx}
\label{table4}
\end{table}

\begin{table}[h]
    \centering
    \caption{Sample Label Distribution (Ternary Classification)}
    \label{tab:sample_label_distribution_ternary}
    \begin{tabular}{c c c}
        \toprule
        Simultaneous Flow Label               & Sample Count & Proportion (\%) \\
        \midrule
        Neither Individual nor Simultaneous Flow  & 278         & 20.1 \\
        Individual but not Simultaneous Flow      & 234         & 17.0 \\
        Simultaneous Flow     & 868         & 62.9 \\
        \bottomrule
    \end{tabular}
    \label{table5}
\end{table}

\subsection{Validation Tools}
Research on flow computations has demonstrated that machine learning and deep learning models are effective in detecing the individual flow experiences \cite{maier2019deepflow}\cite{rissler2020or}\cite{ye2020flow}. Building on these findings, this study employed a range of classic classifiers and neural networks to validate the effectiveness of EEG features for individual and simultaneous flow. To evaluate model performance, this study used metrics such as accuracy, precision, recall and F1 score for a comprehensive evaluation.

\subsubsection{Details of Implementation}
Training and evaluation of all the models were implemented using Python 3.9.12, with the Scikit-Learn\cite{pedregosa2011scikit} library (version 1.2.0) and PyTorch\cite{Paszke_PyTorch_An_Imperative_2019} (version 1.12.0). Model training was performed on Nvidia Geforce RTX 3070 Ti. We used classification models to predict the simultaneous flow experience using the features introduced in Section \labelcref{featureExtraction}. A range of traditional machine learning algorithms from Scikit-Learn were used including Logistic Regression (LR), Support Vector Machines (SVM), Decision Trees (DT), and Random Forests (RF). In addition, we designed several Neural Network(NN) models with different architectures using PyTorch, distinct from Scikit-Learn. Appendix Table \labelcref{tab:model_config} provides a detailed descriptions and configurations of the models.
\subsubsection{Metrics}
For binary and ternary classification tasks, we used metrics including Accuracy, Precision, Recall, and F1 Score. Detailed definitions of these formulas can be found in Appendix Table \labelcref{tab:metrics}.

\subsection{Validation Procedure}
The following strategies were used to reduce the bias introduced by data distribution and the selection of training and testing sets.
\subsubsection{Balancing the Dataset}
For the binary classification task, the ratio of low simultaneous flow to high simultaneous flow samples in the unbalanced dataset was 1:1.64. We used the SMOTE algorithm\cite{chawla2002smote} to generate minority-class data, adjusting the ratio of different classes to 1:1. After adjustment, the sample size for both classes was 838, for a total sample size of 1676.

For the multi-classification task, the ratio of "Neither Individual Flow nor Simultaneous Flow," "Individual but not Simultaneous Flow" and "Simultaneous Flow" in the unbalanced dataset was 1.2:1:3.7. We used the SMOTE algorithm to generate minority class data and adjust the ratio of all three classes to 1:1:1, thereby increasing the total number of samples to 2604.

\subsubsection{Ten-fold Cross-Validation}
To avoid the bias introduced by data partitioning, we used k-fold cross-validation to divide the sample dataset randomly into k folds. In each run, one fold was used as the test set and the remaining k-1 folds were used as the training set. Classifier accuracy, precision, recall, and F1 score were calculated based on the test set for each training set.

To mitigate problems arising from improper data splitting and evaluate the generalizability of the model more accurately, we used ten-fold cross-validation (k=10). The final performance metric was the average of the scores across the ten folds.

\subsection{Validation Results}
\subsubsection{Validation Results of the Binary Classification}
\begin{table}[h!]
\centering
\begin{tabularx}{\textwidth}{lXXXXXXXXXXXXXXXXXX}
    \toprule
    Model & \multicolumn{2}{c}{Accuracy} & \multicolumn{2}{c}{Recall} & \multicolumn{2}{c}{Precision} & \multicolumn{2}{c}{F1 Score} \\
    \midrule
    & Mean & Std & Mean & Std & Mean & Std & Mean & Std \\
    \midrule
            LR & 0.606 & 0.032 & 0.606 & 0.033 & 0.607 & 0.033 & 0.605 & 0.032 \\
		LR (S) & 0.632 & 0.023 & 0.632 & 0.025 & 0.633 & 0.025 & 0.630 & 0.025 \\
		SVM & 0.663 & 0.034 & 0.664 & 0.035 & 0.664 & 0.034 & 0.663 & 0.034 \\
		SVM (S) & 0.696 & 0.030 & 0.697 & 0.032 & 0.698 & 0.033 & 0.696 & 0.030 \\
		DT & 0.618 & 0.049 & 0.618 & 0.049 & 0.620 & 0.050 & 0.616 & 0.049 \\
		DT (S) & 0.757 & 0.039 & 0.757 & 0.039 & 0.757 & 0.039 & 0.756 & 0.039 \\
		RF & 0.758 & 0.032 & 0.759 & 0.033 & 0.760 & 0.033 & 0.758 & 0.032 \\
		RF (S) & \textbf{0.839} & 0.024 & \textbf{0.840} & 0.025 & \textbf{0.841} & 0.024 & \textbf{0.838} & 0.024 \\
		NN & 0.714 & 0.031 & 0.714 & 0.035 & 0.717 & 0.036 & 0.712 & 0.032 \\
		NN (S) & 0.757 & 0.026 & 0.757 & 0.027 & 0.759 & 0.026 & 0.756 & 0.027 \\
		DNN1 & 0.718 & 0.037 & 0.717 & 0.038 & 0.730 & 0.036 & 0.712 & 0.038 \\
		DNN1 (S) & 0.753 & 0.023 & 0.753 & 0.023 & 0.763 & 0.022 & 0.750 & 0.024 \\
		DNN2 & 0.722 & 0.030 & 0.724 & 0.031 & 0.729 & 0.031 & 0.720 & 0.031 \\
		DNN2 (S) & 0.754 & 0.035 & 0.753 & 0.038 & 0.766 & 0.030 & 0.749 & 0.040 \\
		DNN3 & 0.732 & 0.033 & 0.733 & 0.033 & 0.738 & 0.033 & 0.730 & 0.033 \\
		DNN3 (S) & 0.762 & 0.019 & 0.751 & 0.020 & 0.768 & 0.028 & 0.747 & 0.019 \\
    \bottomrule
\end{tabularx}
\caption{Average performance of 10-fold cross validation on binary classification for different machine learning models. Std denotes standard deviation. The configurations of different models are provided in Table \labelcref{tab:model_config} of Appendix.}
\label{table6}
\end{table}
The validation results are introduced as follows. 
Table \labelcref{table6} presents the performance of various machine learning models in a binary classification task, with or without considering inter-brain synchrony features, as evaluated through ten-fold cross-validation. The main results are summarized as follows:

\begin{enumerate}
    \item \textbf{Models including inter-brain synchrony features outperformed those excluding them}, demonstrating their benefits to the classification task. These findings imply that including inter-brain synchrony features can enhance the classification performance of simultaneous flow models, thereby preliminary validating the contribution of the EEG inter-brain synchrony features to simultaneous flow classification. The Appendix's Table \labelcref{tab:p-value-2class} displays the T-test results, showing $p$-values under 0.05 ($ps$<0.05). This suggests the statistical significance of the model's performance enhancement upon including synchrony features.
    \item \textbf{Tree-based models exhibit the most significant improvement after the addition of synchrony features}. When considering synchrony features, the Random Forest model surpassed most others in binary classification, achieving an accuracy of 83.9\%. Decision Trees demonstrated the most notable improvement, enhancing by 13.9\%. This improvement might be attributed to the mechanism of DT and RF. Effective feature extraction and selection ensure the decision nodes split data effectively, thereby enhancing the models’ predictive accuracy. The discussion section provides a more in-depth explanation of the possible reasons for this result.
\end{enumerate}

\subsubsection{Validation Results of the Ternary Classification}

Table \labelcref{table8} presents the performance of various machine learning models in a ternary classification task, considering both with and without inter-brain synchrony features, as evaluated through ten-fold cross-validation. The main results are summarized as follows:

\begin{table}[h]
    \centering
    \begin{tabularx}{\textwidth}{lXXXXXXXXXXXXXXXXXXX}
        \toprule
        Model  & \multicolumn{2}{c}{Accuracy} & \multicolumn{2}{c}{Recall} & \multicolumn{2}{c}{Precision} & \multicolumn{2}{c}{F1 Score}  \\
        \midrule
                         & Mean         & Std       & Mean         & Std       & Mean         & Std       & Mean         & Std       \\
        \midrule
        LR                & 0.607        & 0.029     & 0.607        & 0.029     & 0.604        & 0.029     & 0.604        & 0.03      \\
        LR (S)              & 0.634        & 0.028     & 0.634        & 0.027     & 0.632        & 0.028     & 0.629        & 0.028     \\
        SVM               & 0.729        & 0.020      & 0.730         & 0.022     & 0.729        & 0.022     & 0.725        & 0.020     \\
        SVM (S)             & 0.758        & 0.029     & 0.758        & 0.029     & 0.756        & 0.029     & 0.755        & 0.027     \\
        DT                & 0.629        & 0.022     & 0.629        & 0.023     & 0.628        & 0.022     & 0.625        & 0.022     \\
        DT (S)              & 0.666        & 0.026     & 0.666        & 0.024     & 0.665        & 0.024     & 0.660        & 0.024     \\
        RF                & 0.842        & 0.016     & 0.842        & 0.017     & 0.842        & 0.017     & 0.839        & 0.017     \\
        RF (S)              & 0.868        & 0.019     & 0.870         & 0.019     & 0.869        & 0.019     & 0.867        & 0.019     \\
        NN                & 0.809        & 0.015     & 0.809        & 0.012     & 0.808        & 0.013     & 0.804        & 0.015     \\
        NN (S)              & \textbf{0.872}        & 0.017     & \textbf{0.872}        & 0.014     & 0.874        & 0.017     & \textbf{0.868}        & 0.017     \\
        DNN1              & 0.837        & 0.016     & 0.837        & 0.014     & 0.841        & 0.015     & 0.829        & 0.016     \\
        DNN1 (S)            & 0.862        & 0.020      & 0.863        & 0.015     & 0.868        & 0.019     & 0.856        & 0.019     \\
        DNN2              & 0.843        & 0.020     & 0.842        & 0.023      & 0.846        & 0.021      & 0.836        & 0.023     \\
        DNN2 (S)            & 0.862        & 0.019     & 0.862        & 0.019     & 0.871        & 0.018     & 0.856        & 0.020      \\
        DNN3              & 0.845        & 0.020     & 0.844        & 0.024      & 0.848         & 0.020     & 0.838         & 0.024     \\
        DNN3 (S)            & \textbf{0.872}        & 0.016     & 0.871        & 0.018     & \textbf{0.876}        & 0.019     & 0.866        & 0.018     \\
        \bottomrule
    \end{tabularx}
    \caption{Average performance of 10-fold cross validation on ternary classification for different machine learning models. Std denotes standard deviation. The configurations of different models are provided in Table \labelcref{tab:model_config} of Appendix. }
    \label{table8}  
\end{table}

\begin{enumerate}
    \item \textbf{Consistent with the trends observed in the binary classification task, models including synchrony features generally outperformed those without these features.} This validates the effectiveness of inter-brain synchrony features in enhancing model performance. This suggests that inter-brain synchrony features contain a certain amount of information valuable for the three-class classification. Table \labelcref{tab:p-value-3class} in the Appendix presents the results of the T-test, with $p$-values less than 0.05 ($ps$<0.05), indicating that the improvement in model performance is statistically significant.
    \item \textbf{The metrics of the ternary classification task were generally higher than those of the binary classification task.} This could be attributed to the more distinctive categories in the ternary classification task. The categories in ternary classification task offer clearer distinctions compared to the high and low simultaneous flow states in binary classification. The discussion section provides a more in-depth explanation of the possible reasons for this result.
    \item \textbf{For the performance of different models, the NN and DNN3 with synchrony features outperformed than other models which achieve an accuracy of 87.2\%}. However, we found that more complex DNN models may overfit the training data, especially when the dataset size is inadequate. Overfitting can limit the performance improvement on test data. Compared to the DNN2 with five hidden layers, the DNN3 with nine hidden layers showed limited improvement in performance and offers no advantage in number of parameters. 
\end{enumerate}

\subsubsection{Feature Importance}

After the comparison experiments, we have the following questions:
\begin{itemize}
    \item Q1: Which brain region's features contribute more significantly to flow classification?
    \item Q2: How significant is the contribution of synchrony features to classification?
\end{itemize}

To answer the questions and further validate the effectiveness of synchrony features, we employed feature selection methods using LR, RF, NN and DNN3. Table \labelcref{table:2_syn} and \labelcref{table:3_syn} respectively show the top 20 features for binary and ternary classification tasks including inter-brain synchrony features on four representative models. Tables \labelcref{Feature LR} and \labelcref{Feature RF}, \labelcref{Feature NN} and \labelcref{Feature DNN3} in the Appendix present the top 20 features and their importance values for binary along with ternary classification tasks, including and excluding inter-brain synchrony features. The features and their calculation methods are shown in Table \labelcref{tab:features} of the Appendix.

In the LR model, each feature is associated with a coefficient, indicating the extent and direction of the feature's impact on the prediction. The sign of the coefficient (COEF) denotes the positive or negative correlation of the feature with the result. The magnitude of the coefficient's absolute value reflects the importance of the feature. We calculated the average coefficient for each feature using Logistic Regression models trained in each fold of ten-fold cross-validation. Features were then ranked according to the absolute value of their COEF, selecting the top 20 with the largest absolute values.

In the feature selection based on RF, we used random forest models trained in each fold of ten-fold cross-validation to calculate the average importance of each feature. Features were ranked according to their importance (IMP), and the top 20 features with the highest importance were selected. The importance of a feature was computed based on the Mean Decrease in Impurity (MDI)\cite{breiman2001random}.

SHapley Additive exPlanations (SHAP)\cite{lundberg2017unified} values are a method for interpreting the predictions of machine learning models. SHAP works by assigning the model's output to each input feature, thus showing how much each feature contributes to the prediction. We calculated the average SHAP values for each feature using NN and DNN3 models trained in each fold of ten-fold cross-validation. Features were subsequently ranked based on the absolute values of their SHAP values, and the top 20 features with the largest absolute values were selected.

 \begin{table}[h!]
	\centering
	\begin{tabularx}{\textwidth}{*{4}{>{\centering\arraybackslash}X}}
     \toprule
		\multicolumn{1}{c}{LR}    & \multicolumn{1}{c}{RF}    & \multicolumn{1}{c}{NN}    & \multicolumn{1}{c}{DNN3}  \\
          \midrule
		P7 NFOD                                        & \textbf{P7 DTW $\alpha$}                       & \cellcolor[gray]{0.9}F7 PSD $\alpha$           & \cellcolor[gray]{0.9}\textbf{F7 CCC $\delta$}  \\
		\cellcolor[gray]{0.9}F7 PSD $\alpha$           & \cellcolor[gray]{0.9}\textbf{AF4 CCC $\alpha$} & \cellcolor[gray]{0.9}\textbf{F8 CCC $\delta$}  & \cellcolor[gray]{0.9}AF4 HOZC                  \\
		\cellcolor[gray]{0.9}\textbf{AF4 DTW $\alpha$} & \cellcolor[gray]{0.9}\textbf{F7 CCC $\beta$}   & \cellcolor[gray]{0.9}F7 PSD $\theta$           & \cellcolor[gray]{0.9}\textbf{F3 CCC $\theta$}  \\
		\cellcolor[gray]{0.9}F3 NFOD                   & \cellcolor[gray]{0.9}F4 Mean PSD               & \cellcolor[gray]{0.9}\textbf{AF3 DTW $\delta$} & T7 DE $\delta$            \\
		P7 Mobility                                    & \cellcolor[gray]{0.9}\textbf{F3 CCC $\alpha$}  & \cellcolor[gray]{0.9}\textbf{F8 CCC $\alpha$}  & \cellcolor[gray]{0.9}\textbf{AF4 CCC $\beta$}  \\
		\cellcolor[gray]{0.9}AF3 Mobility              & \cellcolor[gray]{0.9}\textbf{F3 DTW $\theta$}  & \cellcolor[gray]{0.9}F4 LBP $\delta$           & T7 PSD $\beta$            \\
		\cellcolor[gray]{0.9}\textbf{AF3 DTW $\theta$} & \cellcolor[gray]{0.9}\textbf{F7 DTW $\delta$}  & \cellcolor[gray]{0.9}F4 AAFOD                  & \cellcolor[gray]{0.9}F4 LBP $\beta$            \\
		\cellcolor[gray]{0.9}\textbf{F8 DTW $\theta$}  & \cellcolor[gray]{0.9}\textbf{F3 DTW $\alpha$}  & \cellcolor[gray]{0.9}\textbf{F3 DTW $\delta$}  & \cellcolor[gray]{0.9} \textbf{F8 CCC $\delta$}  \\
		T7 PSD $\theta$                                & \cellcolor[gray]{0.9}\textbf{AF3 CCC $\alpha$} & P7 Variance                                    & \cellcolor[gray]{0.9}\textbf{F7 CCC $\alpha$}  \\
		\cellcolor[gray]{0.9}\textbf{F8 DTW $\alpha$}  & \textbf{P7 CCC $\alpha$}                       & \cellcolor[gray]{0.9}\textbf{F4 DTW $\theta$}  & \cellcolor[gray]{0.9}AF4 Energy                \\
		\cellcolor[gray]{0.9}\textbf{F4 DTW $\alpha$}  & \cellcolor[gray]{0.9}\textbf{AF4 CCC $\beta$}  & \cellcolor[gray]{0.9}\textbf{F3 CCC $\theta$}  & \cellcolor[gray]{0.9}F4 HOZC                   \\
		\cellcolor[gray]{0.9}\textbf{F3 DTW $\beta$}   & \cellcolor[gray]{0.9}\textbf{F8 CCC $\alpha$}  & \cellcolor[gray]{0.9}AF3 PPM                   & T7 AAFOD                  \\
		\cellcolor[gray]{0.9}\textbf{AF3 DTW $\alpha$} & \cellcolor[gray]{0.9}\textbf{F8 CCC $\beta$}   & \textbf{P7 DTW $\beta$}                        & \cellcolor[gray]{0.9}AF3 DE FB                 \\
		T7 SD                                          & \cellcolor[gray]{0.9}\textbf{F8 CCC $\delta$}  & T7 PSD $\alpha$                                & \cellcolor[gray]{0.9}F3 DE $\theta$            \\
		T7 Mobility                                    & \cellcolor[gray]{0.9}\textbf{F4 CCC $\alpha$}  & \cellcolor[gray]{0.9}F4 LBP $\theta$           & \cellcolor[gray]{0.9}\textbf{AF3 CCC $\theta$} \\
		\cellcolor[gray]{0.9}\textbf{AF3 DTW $\delta$} & \cellcolor[gray]{0.9}\textbf{AF4 CCC $\theta$} & P7 PSD $\beta$                                 & \cellcolor[gray]{0.9}AF3 PSD $\delta$          \\
		\cellcolor[gray]{0.9}\textbf{F7 DTW $\delta$}  & \cellcolor[gray]{0.9}\textbf{F7 CCC $\delta$}  & \cellcolor[gray]{0.9}\textbf{F8 DTW $\theta$}  & \cellcolor[gray]{0.9}F7 DE $\alpha$            \\
		\cellcolor[gray]{0.9}\textbf{F4 DTW $\beta$}   & \cellcolor[gray]{0.9}F4 average                & \cellcolor[gray]{0.9}F8 PPM                    & T7 Mobility               \\
		\cellcolor[gray]{0.9}F3 Mobility               & \cellcolor[gray]{0.9}F4 PSD $\delta$           & \cellcolor[gray]{0.9}F4 Energy                 & \cellcolor[gray]{0.9}\textbf{AF3 CCC $\beta$}  \\
		T7 PSD $\alpha$                                & \textbf{P7 CCC $\theta$}                       & P7 PPM                                         & \cellcolor[gray]{0.9}F8 Mobility              \\
        \bottomrule
        
	\end{tabularx}
	\caption{The results of feature importance experiment on binary classification models. Features in bold represent inter-brain synchrony features, gray cells represent features extracted from the frontal lobe. The full names of the features are detailed in section \labelcref{featurename}.}
	\label{table:2_syn}
\end{table}

\begin{table}[h]
	\centering
	\begin{tabularx}{\textwidth}{*{4}{>{\centering\arraybackslash}X}}
 \toprule
		\multicolumn{1}{c}{LR}    & \multicolumn{1}{c}{RF}    & \multicolumn{1}{c}{NN}    & \multicolumn{1}{c}{DNN3}  \\
  \midrule
		P7 SD                                          & \cellcolor[gray]{0.9}F4 PSD $\theta$           & P7 PSD $\beta$                                 & \cellcolor[gray]{0.9}\textbf{AF4 CCC $\theta$} \\
		\cellcolor[gray]{0.9}\textbf{AF4 DTW $\alpha$} & P7 \textbf{DTW $\alpha$}                       & \cellcolor[gray]{0.9}F7 PSD $\beta$            & P7 Kurtosis               \\
		\cellcolor[gray]{0.9}AF4 Mobility              & \cellcolor[gray]{0.9}F4 average                & \cellcolor[gray]{0.9}\textbf{F8 DTW $\delta$}  & \cellcolor[gray]{0.9}AF3 Kurtosis              \\
		\cellcolor[gray]{0.9}F4 DE $\delta$            & \cellcolor[gray]{0.9}AF3 Kurtosis              & P7 Kurtosis                                    & T7 LBP $\beta$            \\
		P7 Mobility                                    & \cellcolor[gray]{0.9}AF3 PSD $\beta$           & \cellcolor[gray]{0.9}F8 PSD $\theta$           & \cellcolor[gray]{0.9}F4 \textbf{CCC $\theta$}  \\
		\cellcolor[gray]{0.9}\textbf{AF3 DTW $\theta$} & P7 DE $\theta$                                 & \cellcolor[gray]{0.9}AF4 PSD $\delta$          & \cellcolor[gray]{0.9}F7 DE $\beta$             \\
		\cellcolor[gray]{0.9}F7 NFOD                   & \cellcolor[gray]{0.9}F8 PPM                    & \cellcolor[gray]{0.9}AF3 LBP $\beta$           & \cellcolor[gray]{0.9}F7 DE FB                  \\
		\cellcolor[gray]{0.9}F8 SD                     & \cellcolor[gray]{0.9}F4 LBP $\delta$           & \cellcolor[gray]{0.9}AF3 PSD $\beta$           & \cellcolor[gray]{0.9}F7 Kurtosis               \\
		P7 Variance                                    & P7 LBP $\theta$                                & \cellcolor[gray]{0.9}AF4 AAFOD                 & \cellcolor[gray]{0.9}F4 LBP $\beta$            \\
		P7 Activity                                    & \cellcolor[gray]{0.9}F4 Energy                 & \cellcolor[gray]{0.9}F4 DE $\delta$            & \cellcolor[gray]{0.9}AF3 \cellcolor[gray]{0.9}DE $\theta$           \\
		T7 DE FB                                       & \cellcolor[gray]{0.9}F8 Kurtosis               & P7 HOZC                                        & \cellcolor[gray]{0.9}F3 NOFD                   \\
		\cellcolor[gray]{0.9}AF4 DE FB                 & \cellcolor[gray]{0.9}F4 Power                  & \cellcolor[gray]{0.9}\textbf{F4 DTW $\delta$}  & \textbf{T7 CCC $\theta$}  \\
		P7 NFOD                                        & \textbf{P7 CCC $\beta$}                        & \cellcolor[gray]{0.9}F3 LBP $\theta$           & \cellcolor[gray]{0.9}P7 DE FB                  \\
		\cellcolor[gray]{0.9}F8 DE FB                  & \cellcolor[gray]{0.9}\textbf{AF4 CCC $\alpha$} & T7 PSD $\theta$                                & \cellcolor[gray]{0.9}F4 LBP $\theta$           \\
		\cellcolor[gray]{0.9}AF4 PPM                   & \cellcolor[gray]{0.9}\textbf{F8 CCC $\alpha$}  & T7 DE FB                                       & \textbf{T7 CCC $\delta$}  \\
		\textbf{T7 DTW $\delta$}                       & \cellcolor[gray]{0.9}F7 PSD $\beta$            & \cellcolor[gray]{0.9}F3 DE $\theta$            & \cellcolor[gray]{0.9}\textbf{F3 CCC $\alpha$}  \\
		\cellcolor[gray]{0.9}F7 Mobility               & \cellcolor[gray]{0.9}F4 LBP $\alpha$           & T7 PSD $\alpha$                                & \cellcolor[gray]{0.9}\textbf{F4 DTW $\delta$}  \\
		\cellcolor[gray]{0.9}F8 PSD $\beta$            & \textbf{T7 DTW $\theta$}                       & \cellcolor[gray]{0.9}\textbf{AF3 DTW $\alpha$} & \cellcolor[gray]{0.9}\textbf{F7 CCC $\delta$}  \\
		\cellcolor[gray]{0.9}\textbf{F4 DTW $\alpha$}  & \cellcolor[gray]{0.9}F4 DE $\alpha$            & P7 PSD $\alpha$                                & T7 LBP $\theta$           \\
		\cellcolor[gray]{0.9}AF4 DE $\delta$           & \textbf{T7 DTW $\beta$}                        & \cellcolor[gray]{0.9}\textbf{F7 CCC $\delta$}  & T7 AAFOD  \\
            \bottomrule
	\end{tabularx}
	\caption{The results of feature importance experiment on ternary classification models. Features in bold represent inter-brain synchrony features, gray cells represent features extracted from the frontal lobe. The full names of the features are detailed in section \labelcref{featurename}.}
	\label{table:3_syn}
\end{table}

%


The main results are summarized as follows:
\begin{enumerate}
    \item \textbf{Compared to features of the left temporal lobe (T7, P7), the features located in the frontal lobe (AF3, AF4, F3, F4, F7, F8) were more preferred by the models.} As shown in Tables \labelcref{table:2_syn} and \labelcref{table:3_syn}, the frontal lobe features (gray cells) account for a significant proportion both in binary and ternary classification. This suggests that the frontal lobe region likely carries more effective information than the left temporal lobe. This is consistent with Bruya et al.'s opinion\cite{bruya2010effortless} that the flow experience is closely related to the activation level of cerebral cortex and frontal cortex.
    \item \textbf{The inclusion of synchrony features accounted for a significant proportion in the top 20 features.} In the binary classification task including synchrony features, the synchrony features occupied over 35\% of the top 20 features, while in ternary classification tasks, they represented over 20\%. This substantiates the efficacy of inter-brain synchrony features in recognizing simultaneous flow.
    \item \textbf{Synchrony features were more focused in binary classification tasks.} The preference for synchrony features in binary classification may attribute to their critical role in reflecting team collaboration levels to differentiate between high and low simultaneous flow. In contrast, the ternary classification requires a balance between individual and synchrony features due to its clearer category distinctions. The discussion section provides a more in-depth explanation of the possible reasons for this result.

\end{enumerate}

\subsubsection{Ablation Study}
To evaluate the impact of different feature sets on the performance of different models, we conducted ablation experiment. This experiment aims to further validate the significance of features related to the frontal lobe and inter-brain synchrony. The models considered include LR, RF, NN and DNN3. Each model was trained using a 10-fold cross-validation to ensure the reliability of the results. Table \labelcref{table11} shows the results of ablation experiments for LR, RF, NN, DNN3 on ternary classification.

The feature sets used are extracted from the left temporal lobe (L), the frontal lobe (F), and their corresponding synchrony features (LS for left temporal lobe, FS for frontal lobe). The baseline feature set consisted of features from the left temporal lobe (L). We calculated $\Delta$ to represent the percentage increase in performance metrics.

\begin{table}[h]
    \centering
    \begin{tabularx}{\textwidth}{l|l|*{8}{>{\centering\arraybackslash}X}}
        \toprule
        Model  & \multicolumn{1}{c}{Feature Set} & \multicolumn{2}{c}{Accuracy} & \multicolumn{2}{c}{Recall} & \multicolumn{2}{c}{Precision} & \multicolumn{2}{c}{F1 Score}  \\
        \midrule
                                &                             & Mean           & $\Delta$(\%) $\uparrow$      & Mean          &$\Delta$(\%) $\uparrow$        & Mean           & $\Delta$(\%) $\uparrow$       & Mean           & $\Delta$(\%) $\uparrow$        \\
        \midrule
            \multirow{6}{*}{LR} & L         & 0.479 & -    & 0.480 & -    & 0.478 & -    & 0.476 & -    \\
		                      & F         & 0.580 & 10.1 & 0.580 & 10.0 & 0.576 & 9.8  & 0.574 & 9.8  \\
		                      & L+F       & 0.607 & 12.8 & 0.607 & 12.7 & 0.604 & 12.6 & 0.604 & 12.8 \\
		                      & L+F+LS    & 0.608 & 12.9 & 0.608 & 12.8 & 0.604 & 12.6 & 0.603 & 12.7 \\
		                      & L+F+FS    & 0.633 & 15.4 & 0.633 & 15.3 & 0.630 & 15.2 & 0.629 & 15.3 \\           
		                      & L+F+FS+LS & 0.634 & 15.5 & 0.634 & 15.4 & 0.632 & 15.4 & 0.629 & 15.3 \\
            \midrule
		\multirow{6}{*}{RF} & L         & 0.772 & -    & 0.772 & -    & 0.772 & -    & 0.768 & -    \\
		                      & F         & 0.831 & 5.9  & 0.831 & 5.9  & 0.830 & 5.8  & 0.828 & 6.0  \\
		                      & L+F       & 0.842 & 7.0  & 0.842 & 7.0  & 0.842 & 7.0  & 0.839 & 7.1  \\
		                      & L+F+LS    & 0.851 & 7.9  & 0.852 & 8.0  & 0.851 & 7.9  & 0.850 & 8.2  \\
		                      & L+F+FS    & 0.856 & 8.4  & 0.857 & 8.5  & 0.857 & 8.5  & 0.855 & 8.7  \\           
		                      & L+F+FS+LS & 0.868 & 9.6  & 0.870 & 9.8  & 0.869 & 9.7  & 0.867 & 9.9  \\
            \midrule
		\multirow{6}{*}{NN} & L         & 0.630 & -    & 0.631 & -    & 0.631 & -    & 0.624 & -    \\
		                      & F         & 0.796 & 16.6 & 0.796 & 16.5 & 0.795 & 16.4 & 0.791 & 16.7 \\
		                      & L+F       & 0.809 & 17.9 & 0.809 & 17.8 & 0.808 & 17.7 & 0.804 & 18.0 \\
		                      & L+F+LS    & 0.845 & 21.5 & 0.846 & 21.5 & 0.847 & 21.6 & 0.840 & 21.6 \\
		                      & L+F+FS    & 0.863 & 23.3 & 0.864 & 23.3 & 0.865 & 23.4 & 0.860 & 23.6 \\
		                      & L+F+FS+LS & 0.872 & 24.2 & 0.872 & 24.1 & 0.874 & 24.3 & 0.868 & 24.4 \\
            \midrule
	   \multirow{6}{*}{DNN3} & L         & 0.762 & -    & 0.762 & -    & 0.766 & -    & 0.753 & -    \\
		                      & F         & 0.824 & 6.2  & 0.823 & 6.1  & 0.831 & 6.5  & 0.815 & 6.2  \\
		                      & L+F       & 0.845 & 8.3  & 0.844 & 8.2  & 0.848 & 8.2  & 0.838 & 8.5  \\
		                      & L+F+LS    & 0.858 & 9.6  & 0.858 & 9.6  & 0.864 & 9.8  & 0.852 & 9.9  \\
                                & L+F+FS    & 0.868 & 10.6 & 0.868 & 10.6 & 0.871 & 10.5 & 0.863 & 11.0 \\
		                      & L+F+FS+LS & 0.872 & 11.0 & 0.871 & 10.9 & 0.876 & 11.0 & 0.866 & 11.3 \\
        \bottomrule
    \end{tabularx}
    \caption{Ablation experiments for LR, RF, NN, DNN3 on ternary classification. $\Delta$ represents the percentage increase in performance metrics compared to the baseline feature set (using only left temporal lobe features, denoted as 'L'). L stands for features from the left temporal lobe, F for features from the frontal lobe, FS for synchrony features of the frontal lobe, and LS for synchrony features of the left temporal lobe. The same model configurations are used across different feature sets, as detailed in Table \labelcref{tab:model_config} of Appendix.}
    \label{table11}  
\end{table}

The main results are summarized as follows:
\begin{enumerate}
    \item \textbf{Feature addition improves performance and the improvements across different metrics is consistent.} As shown in Table \labelcref{table11}, the inclusion of additional features generally improved the performance metrics for all models. Moreover, the consistent trend of performance improvements across all metrics for each model indicates the robustness of performance gains.
    \item \textbf{Models using frontal lobe features(F) exhibit significant improvements on all metrics compared to which uses using left lobe features(L)}. The performance enhancements($\Delta$) across four metrics range from 5.8\% to 16.7\% which suggest features from frontal lobe carry more information related to flow experience.
    \item \textbf{Synchrony features in the frontal lobe appear to be more significant than those in the left temporal lobe.} The models exhibited greater performance enhancements with the inclusion of the FS feature set compared to the LS feature set. This suggests that synchrony features from the frontal lobe contain more relevant information, contributing to greater performance gains compared to those from the left temporal lobe.
    \item \textbf{Inter-brain synchrony features played a significant role in classifying flow states during cooperative tasks.} Performance improvements were observed in nearly all models following the addition of the LS, FS, and combined LS+FS feature sets. This result, in conjunction with the results of the feature importance experiments, collectively validates the importance of inter-brain synchrony features.
\end{enumerate}

\section{Discussion}
This study focus on the simultaneous flow, preliminarily exploring its features and the methods for detecting simultaneous flow experiences in collaboration based on EEG signals. The following discussion addresses several important questions.

\subsection{Validity of Simultaneous Flow Tasks and Dataset}
\subsubsection{The two-player Whack-A-Mole game is effective in inducing simultaneous flow experiences.} Experiments on neural synchrony in simultaneous flow are rare. And There is no unified task paradigm is specifically designed to induce simultaneous flow. Our study used a two-player Whack-A-Mole game, which is commonly used in individual flow research. The task is rule-based and straightforward, allowing for quick concentration and the induction of flow experiences in a short time. In addition, the task is robust to isolated and random operator errors and is well-suited to traditional flow evaluation techniques, such as the Experience Sampling Method(ESM). Our results indicate that the experimental task is effective in inducing simultaneous flow experiences in the participants.
\subsubsection{The EEG dataset to classify simultaneous flow was constructed effectively.} Rather than aiming for the highest number of channels, we selected those with most potential relevance to individual and simultaneous flow. Based on the neural mechanisms of flow, the frontal lobe brain regions are highly associated with flow experiences. In addition, the left temporal lobe, which is involved in advanced neural synchrony, was preliminarily found activated during flow in collaborative tasks (relevant to simultaneous flow). Therefore, we focused on the EEG signals from the frontal (AF3, AF4, F3, F4, F7, F8) and left temporal (T7, P7) lobes to construct the EEG dataset of simultaneous flow. 

All previous dataset of flow focused on individual flow, labeling types of experience induced by different match of skills and challenges, such as boredom, flow and anxiety\cite{katahira2018eeg}\cite{plotnikov2012exploiting}\cite{sinha2015dynamic}\cite{ye2020flow}. In contrast, our study classified experiences into high and low simultaneous flow, as well as situations that were "Neither Individual nor Simultaneous Flow", "Individual but not Simultaneous Flow", and "Simultaneous Flow". Not only did we delve into the intensity of simultaneous flow, but also considered on the relationship between individual and simultaneous flow, and our results show that this EEG dataset contains effective flow labels and features suitable for detecting simultaneous flow experiences.

\subsection{EEG Features of Simultaneous Flow}
\subsubsection{It is necessary to focus the features that reflect the synchrony of neural activity}
According to Yun et al.\cite{Yun_Watanabe_Shimojo_2012}, the synchrony of neural activity between the participants increased after cooperative interaction. Sinha et al.\cite{sinha2016eeg} have demonstrated that inter-brain synchrony is more likely to occur when participants are cooperating towards achieving a common goal. Simultaneous flow is a unique brain experience associated with enhanced information integration and neural synchrony. Therefore, to assess simultaneous flow, it is necessary to focus not only on relevant brain regions but also on EEG features that reflect the synchrony of neural activity across the brain. The global inter-brain integrated information and neural synchrony are enhanced in the state of team flow(relevant to simultaneous flow), particularly with specific fluctuations observed in the left temporal lobe cortex\cite{shehata2020team}. The individual flow experience is closely related to the activation level of the frontal lobe\cite{bruya2010effortless}. As scientific knowledge of the neural mechanisms of simultaneous flow is scarce, our study initially used the cross-correlation coefficient to measure the signal correlation between participants and Dynamic Time Warping (DTW) to quantify the similarity between two signals. Although these features may not fully capture all aspects of inter-brain synchrony, they serve as a first attempt.

\subsubsection{The inter-brain synchrony features we extracted is effective}
To validate the effectiveness of these features, we employed eight commonly used classifiers and neural networks. Our comparison experiments across binary and ternary classification tasks demonstrated that the inclusion of the preliminary inter-brain synchrony features improved the performance of all eight models significantly. Furthermore, synchrony features were prominently represented among the top-20 features in feature importance experiments, suggesting their importance. Notably, performance enhancements were observed in nearly all models with the addition of the LS, FS, and combined LS+FS feature sets. These results collectively validated the effectiveness of inter-brain synchrony features in detecting simultaneous flow.
\subsection{Classification of Simultaneous Flow}
\subsubsection{The simultaneous flow computation pipeline proposed in this paper is feasible.} We employed a ten-fold cross-validation method to evaluate the performance of LR, SVM, DT, RF, NN, DNN1, DNN2, and DNN3 in binary and ternary classification. The accuracy of these machine learning models surpassed the results of the random selection (50\%, 33.3\%). This suggests the feasibility of the simultaneous flow computation pipeline proposed in this study, including data acquisition, feature extraction, and the selection of machine learning models. We validated the effectiveness of inter-brain synchrony features in simultaneous flow computation for the first time, offering a novel perspective for this field. 
\subsubsection{Random Forest exhibited the best performance in binary classification task}
In binary classification tasks, the Random Forest model including inter-brain synchrony features exhibited the best performance with an accuracy of 83.9\%, surpassing DNN3 by 8.3\%. More complicated models may encounter the problem of overfitting with limited user data. If a larger data sample could be collected as training data, the DNN models could potentially achieve better performance.
\subsubsection{DT and RF showed significant improvements in experiments when including synchrony features on binary classification ($ps$ < 0.001).} 
\label{dis:tree_up}
Decision Trees improved by 13.9\%, while Random Forest enhanced by 8.1\%. Neural network-based models like NN and DNN already perform well without the addition of synchrony features, but their improvement is not as significant as tree models when these features are included. This might be because neural networks usually need to encode categorical features into numerical forms to accommodate the mathematical operations of the network. Such transformations can lead to information loss, whereas tree models inherently support the processing of categorical features. Tree models create decision rules by splitting data, forming different paths for each feature's distinct values. This is also evidenced in Table \labelcref{table:2_syn} and \labelcref{table:3_syn}, where RF focused more on the synchrony features.

\subsubsection{Ternary classification metrics generally surpass binary ones.}
Ternary classification metrics typically surpassed those of binary classification, This could be attributed to the more distinctive categories in the ternary classification task. In the ternary classification task, the categories are defined as follows: 1) Neither Individual nor Simultaneous Flow; 2) Individual but No Simultaneous Flow; 3) Simultaneous Flow. These categories offer clearer distinctions compared to the high and low simultaneous flow states in binary classification. Specifically, the category of "individual flow but no simultaneous flow" cannot be effectively identified in binary classification, and might be mistakenly classified as low or high simultaneous flow. As shown in Table \labelcref{table:3_syn}, due to the clearer category distinctions, the ternary classification requires a balance between individual and synchrony features.

\subsubsection{The features located in the frontal lobe are more preferred by the models.} Compared to the features of the left temporal lobe (T7, P7), the features located in the frontal lobe (AF3, AF4, F3, F4, F7, F8) are more preferred by the models. However, this observation slightly contrasts with the theory proposed by Shehata et al.\cite{shehata2020team}, which suggests that the activation of the left temporal cortex during team flow experiences. This difference can be attributed to task designs: Shehata et al. conducted their tasks on the same iPad, whereas our study designed different tasks performed on two separate computers. In our experiment, both participants exhibited high levels of concentration, the degree of individual flow is stronger than that of team flow. This aligns with the opinion of Bruya et al.\cite{bruya2010effortless} that "individual flow is closely related to the activation level of the frontal lobe."
\subsubsection{From simultaneous flow to team flow}
Simultaneous flow means team members experiencing individual flow at the same time. Van den Hout et al. define team flow as a concatenative experience of team members' individual flow. In our experiments, both participants aimed to achieve the common team goal and optimize team performance, they exhibited high levels of concentration in the state of simultaneous flow. When each participant is in individual flow and doing their own tasks for the team performance, it might lead to the shared team flow together.

\subsection{Limitations and Future Directions}
There remain some limitations to this study that need to be addressed in future studies.

First, further exploration of simultaneous flow features reflecting inter-brain synchrony. This study extracted features related to inter-brain synchrony in team flow. However, as a first attempt, these features were relatively simple. Future work could explore a more diverse set of features such as coherence and phase synchrony. Therefore, ongoing research on the neural mechanisms of simultaneous flow is required to provide a more robust neuroscientific basis for feature extraction.

Second, extending simultaneous flow induction tasks and datasets to larger teams. As an exploratory study, this research began with the smallest scale dyadic teams to construct flow tasks and EEG datasets. In the future, higher precision EEG devices will be used to acquire signals and it will be important to extend these tasks and datasets to larger teams to accommodate a wider variety of team-interaction scenarios.

Third, real-time detection of simultaneous flow experience. This study focused on EEG-based simultaneous-flow detection methods. Through feature extraction and training machine learning models, we achieved the detection of simultaneous flow experience. However, this method does not currently provide a real-time evaluation of flow experience. Based on this study, future work could be expected to achieve a real-time detection of simultaneous flow experiences for evaluation and human-computer interaction purposes.

Finally, experiencing flow simultaneously is foundational to achieving shared team flow. Simultaneous flow is an important step before focusing on team flow. Future work will explore how to induce and detect team flow in different tasks, potentially enhancing team performance in multi-user collaborative systems.

\section{Conclusion}
This study explored methods for detecting simultaneous flow based on EEG signals. The main conclusions of this study are as follows.

First, we designed a two-user collaborative experimental task to effectively induce simultaneous flow. Based on this task, we recorded the user’s eight-channel EEG signals (F3, F4, F7, F8, AF3, AF4, T7, and P7) from the frontal lobe brain regions (related to individual flow) and left temporal lobe brain regions (related to inter-brain synchrony of team flow) and then constructed the first effective EEG dataset with multiple labels of simultaneous flow.


Second, we extracted specific features relevant to individual flow and inter-brain synchrony, and proved that using them can effectively detecting different types of simultaneous flow. Particularly, We preliminarily proposed inter-brain synchrony features relevant to simultaneous flow from EEG signals for the first time, including Cross-Correlation Coefficients and Dynamic Time Warping Distance. Based on various machine learning models, we validated their effectiveness in enhancing the machine learning models' performance in binary and ternary classification tasks.


Third, We compared the metrics on different models and conducted a thorough analysis of the experiments. We ranked the extracted features in terms of importance by using LR, RF, NN and DNN3 and conducted a ablation study. Based on the results of these experiments, we found the inter-brain synchrony features has high importance in simultaneous flow detection and the features from frontal lobe area are given priority attention by the models.


\bibliographystyle{ACM-Reference-Format}
\bibliography{manuscript}

\appendix
\setcounter{table}{0}

\section{Appendix}
\subsection{Metrics}

\begin{longtable}[c]{p{0.15\textwidth}p{0.35\textwidth}p{0.5\textwidth}}
    \toprule
    Metrics & Description & Definition \\
    \hline
    \endfirsthead
    \toprule
    Metrics & Description & Definition \\
    \hline
    \endhead
    \hline
    \endfoot
    \hline
    \caption{Evaluation Metrics and Calculation Methods}
    \label{tab:metrics}
    \endlastfoot
    \multicolumn{3}{c}{Binary Classification Metrics} \\
    \midrule
		Accuracy  & The proportion of true results among the total number of cases examined. & $\text{Accuracy} = \frac{TP + TN}{TP + TN + FP + FN} $   \\
  \midrule
		Precision & The proportion of positive identifications that were actually correct. & $ \text{Precision} = \frac{TP}{TP + FP} $                 \\
  \midrule
		Recall    & The proportion of actual positives that were correctly identified. & $ \text{Recall} = \frac{TP}{TP + FN} $                    \\
  \midrule
		F1 Score  & The harmonic mean of precision and recall, providing a balance between them. & $\text{F1 Score} = 2 \times \frac{Pre \times Rec}{Pre + Rec}$   \\
  \midrule
  \multicolumn{3}{c}{Ternary Classification Metrics} \\
    \midrule
		Accuracy  & The proportion of true results for a three-class classification problem. & $\text{Accuracy} = \frac{\sum_{i=1}^{3}TP_i + \sum_{i=1}^{3}TN_i}{N}$                                                          \\
  \midrule
		Precision & The average proportion of correct positive predictions for each class. & $\text{Precision} = \frac{1}{3}\left(\frac{TP_1}{TP_1+FP_1} + \frac{TP_2}{TP_2+FP_2} + \frac{TP_3}{TP_3+FP_3}\right)$                       \\
  \midrule
		Recall    & The average proportion of correctly identified positives for each class. & $\text{Recall} = \frac{1}{3}\left(\frac{TP_1}{TP_1+FN_1} + \frac{TP_2}{TP_2+FN_2} + \frac{TP_3}{TP_3+FN_3}\right)$                          \\
  \midrule
		F1 Score  & The harmonic mean of precision and recall for ternary classification. & $\text{F1 Score} = 2 \times \frac{Pre \times Rec}{Pre + Rec}$ \\
  \bottomrule
\end{longtable}

\( TP \) is the number of true positives.
\( FP \) is the number of false positives.
\( TN \) is the number of true negatives.
\( FN \) is the number of false negatives.

\subsection{Features}

\begin{longtable}[c]{p{0.35\textwidth}p{0.3\textwidth}p{0.35\textwidth}}
    \toprule
    Feature Category & Description & Definition \\
    \hline
    \endfirsthead
    \toprule
    Feature Category & Description & Definition \\
    \hline
    \endhead
    \hline
    \endfoot
    \hline
    \caption{Features Extracted for Individual and Team Flow}
    \label{tab:features}
    \endlastfoot
	Mean                                                                      & The mean as a feature of the EEG signal reflects the average level of electrical potential over a period of time, revealing the baseline state of brain activity.                                                                                                                       & $\mu_s = \frac{1}{T} \sum_{n=1}^{T} s(n)$                                                                                                                                   \\
 \midrule
	Standard Deviation                                                        & The standard deviation as a feature of the EEG signal indicates the dispersion of signal values, quantifying the variability or stability of brain activity.                                                                                                                       & $\sigma_s = \sqrt{\frac{1}{T} \sum_{n=1}^{T} (s(n) - \mu_s)^2}$                                                                                                                \\
 \midrule
  Variance\cite{zhang2019review} & The variance represents the range of fluctuation of the EEG signal and also the power of the signal to deviate from the mean value. & $\text{Var}_s = \frac{1}{T} \sum_{n=1}^{T} (s(n) - \mu_s)^2$ \\
 \midrule
	Average Absolute \newline First-Order Difference\cite{zhang2019review}             & The average absolute value of the first-order difference measures the rate of change of the EEG signal over time, helping to capture short-term fluctuations within the signal.                                                                                                                       & $\delta_s = \frac{1}{T-1} \sum_{n=1}^{T-1} \left| s(n+1) - s(n) \right|$                                                                                                    \\
 \midrule
	Normalized First-Order Difference\cite{zhang2019review}                   &  The normalized first-order difference eliminates the influence of signal strength, making the features more comparable under different conditions.                                                                                                                      & $\overline{\delta_s} = \frac{\delta_s}{\sigma_s}$                                                                                                                           \\
 \midrule
	Energy\cite{hernandez2014using}                                           & 
Fluctuations in energy can reflect the level of activity in the cerebral cortex, often represented by the square of the time-domain signal.                                                                                                                       & $E_s = \sum_{n=1}^{T} s(n)^2$                                                                                                                                               \\
 \midrule
	Power\cite{zhang2019review}                                               & 
The power of the EEG signal is obtained by dividing the energy by the number of samples.                                                                                                                      & $P_s = \frac{1}{T} \sum_{n=1}^{T} s(n)^2$                                                                                                                                   \\
 \midrule
	Hjorth Parameters\cite{sinha2015dynamic}                                  & Activity measures the deviation of the amplitude, Mobility measures the change in slope.                                & $\text{Activity} = \sigma_s^2$ \newline $\text{Mobility} = \frac{\sigma_f}{\sigma_s}$                                                                                                \\
 \midrule
	Higher-Order Zero-Crossing\cite{petrantonakis2009emotion,zhang2019review} & Higher-order zero-crossing reflects the oscillation level of the EEG signal, indicating the fluctuation features.       & By using the mean $ \mu_s $, transform the EEG signal into a zero-mean sequence $ s_{\text{new}}(t)=s(t)-\mu_s $. \newline $ H_s = \sum_{n=1}^{T-1} (x_s(t+1) - x_s(t))^2 $ \\
 \midrule
	Peak-to-Peak Mean\cite{li2018exploring}                                   & Peak-to-peak mean is the arithmetic mean of the distance between the maximum and minimum values in a time-series signal. & $\text{ppm}_s = \frac{\max_{t_i \in t} s(t_i) - \min_{t_i \in t} s(t_i)}{T}$                                                                                                \\
 \midrule
	Kurtosis\cite{alsharabi2022eeg,wagh2022performance}                       & Kurtosis reflects the frequency of peak values appearing as well as the waveform features of the signal.                & $\text{Kur}_s = \frac{E(s(t) - \mu_s)^4}{\sigma_s^4}$                                                                                                                       \\
 \midrule
	Logarithmic Band Power\cite{10109720}                                     & 
The logarithmic band power converts the frequency information of EEG signals into an energy distribution on a logarithmic scale.                                                                                                                      & $\text{LBP}_s = \log\left(\frac{1}{T} \sum_{n=1}^{T} s(n)^2\right)$                                                                                                         \\
 \midrule
	Differential Entropy\cite{duan2013differential,10109720}                  & 
Differential entropy is often used to represent the complexity of a time series.                                                                                                                       & $\text{DE}_s = \frac{1}{2} \log\left(2\pi e \sigma_s^2\right)$                                                                                                              \\
 \midrule
	Power Spectral Density\cite{alazrai2018eeg,10109720}                      & The power spectral density describes how the power of a signal varies with frequency.                                                                                                                       & $\text{PSD}_s = \frac{1}{2} \log\left(2\pi e \sigma_s^2\right)$                                                                                                             \\
 \midrule
	Cross-Correlation Coefficient                                             & The Cross-Correlation Coefficient is used to measure the similarity between two time series at different time lags                                                                                                                       & $\rho_{A,B} = \frac{\text{cov}(A, B)}{\sigma_A \sigma_B}$                                                                                                                   \\
 \midrule
	Dynamic Time Warping Distance                                             &  The Dynamic Time Warping Distance is used to measure the similarity between two time series.                                                                                                                       & $D(a_i, b_j) = \text{Dist}(a_i, b_j) + \min\{D(a_{i-1}, b_j), D(a_i, b_{j-1}), D(a_{i-1}, b_{j-1})\}$                                                                       \\

\end{longtable}

\subsection{Details of Implementation}
\begin{longtable}{p{0.35\textwidth}p{0.35\textwidth}p{0.3\textwidth}}
    \toprule
    Model & Description & Configuration \\
    \hline
    \endfirsthead
    \toprule
    Model & Description & Configuration \\
    \hline
    \endhead
    \hline
    \endfoot
    \hline
    \caption{Machine learning models and configurations}
    \label{tab:model_config}
    \endlastfoot    
	Logistic Regression(LR)                & Logistic Regression uses a linear function of input features $ x $ to predict the label $y$. The label $y$ can be either binary or multi-class. In binary classification, the model predicts the probability of the label belonging to one of two categories. This is usually represented as $\hat{y} = \frac{1}{1 + e^{-w^T x}}$, where $ e $ is the base of the natural logarithm, and $w$ represents the model parameters. In multi-class scenarios, logistic regression typically uses the softmax function to handle multiple categories and predict the probability of each category. The goal of the model is to estimate parameters \(w\) by maximizing the likelihood function, usually achieved by minimizing the log loss of the training data. & Feature standardization was performed using the StandardScaler parameter, followed by the sklearn's LogisticRegression model. The maximum number of iterations, max\_iter, was set to 1000. For the ternary classification task, we specified multi\_class='multinomial', enabling the model to handle multi-class problems.                                                                                                                                                                                                         \\
     \midrule
	Support Vector Machine (SVM)           & SVM classification models predict the label $ \hat{y} = w^T x + b $ by solving the following optimization problem:$ \min_{w} \frac{1}{2} \| w \|^2 $ subject to the classification constraint:$ y_{\text{train}}(w^T x_{\text{train}} + b) \geq 1 - \xi $ where $ x $ represents the input features, $ w $ is the weight vector, $ b $ is the bias term, and $ \xi $ are slack variables allowing for some data points to violate the margin. For binary classification tasks, the model output $ \hat{y} $ is based on the sign function determining to which category a data point belongs. In ternary or multi-class scenarios, SVM can be extended to multi-category classification by using one-vs-rest or one-vs-one strategies.& The SVM model is configured with C=1.0, kernel='rbf', and gamma='scale'. The C parameter controls the penalty for misclassification, and kernel specifies the kernel type. The value 'scale' means that gamma is set to $\frac{1}{ n_{features} * X.var()}$, where X is the feature data. This setting takes into account the variance of features and automatically adjusts the gamma value.                                                                                                                                      \\
 \midrule
	Decision Tree (DT)                     & The Decision Tree model we used is based on Classfication and Regression Tree (CART)\cite{breiman2017classification}. A binary tree is constructed by iteratively splitting data based on feature thresholds that maximize class separation.                                                                                                                                                                                                                                                                                                                                                                                                                                                                                                                                                & The Decision Tree model is configured with  min\_samples\_split=4.  The min\_samples\_split parameter specifies the minimum number of samples required to split an internal node. The splitting stops when the number of samples is less than min\_samples\_split.                                                                                                                                                                                                                                                                       \\
 \midrule
	Random Forest (RF)                     & Random Forest\cite{breiman2001random} is an ensemble of Decision Trees. It trains each tree on a random subset of the training data and uses a random subset of features for each split. The final prediction is the average prediction of all trees.                                                                                                                                                                                                                                                                                                                                                                                                                                                                                                                               & The Random Forest model is configured with max\_features=0.3, bootstrap=ture, max\_samples=0.8 and min\_samples\_leaf=3. The max\_features parameter specifies the number of features to consider when looking for the best split. The bootstrap parameter specifies whether bootstrap samples are used when building trees. The max\_samples parameter specifies the number of samples to draw from X to train each base estimator. The min\_samples\_leaf parameter specifies the minimum number of samples required to be at a leaf node. \\
 \midrule
	Neural Network (NN) (1-hidden)         & Neural Network\cite{rumelhart1986learning} is a model inspired by the biological neural networks. It consists of multiple layers of neurons. Each neuron is a linear function of the input features followed by a non-linear activation function. The output of the network is the output of the last layer.The output layer consists of an affine transform to predict the flow experience.                                                                                                                                                                                                                                                                                                                                                                                            & The NN includes a single hidden layer with 100 neurons, using ReLU function\cite{nair2010rectified}. Batch normalization\cite{ioffe2015batch} is applied to the inputs of each layer, and dropout\cite{srivastava2014dropout} is set at a rate of 0.1. The network is optimized using the Adam optimizer\cite{Kingma_Ba_2014} with a learning rate of 0.001, and it employs cross-entropy loss as the loss function. We set the batch size to 64 for 10-folds cross validation.                                                                                                                 \\
 \midrule
	Deep Neural Network1 (DNN1) (5-hidden) & The DNN refers to Neural Network with multiple hidden layers\cite{Hinton_Osindero_Teh_2006}.                                                                                                                                                                                                                                                                                                                                                                                                                                                                                                                                                                                                                                                                                               & In DNN1, we use 5 hidden layers where each layer consists of a batch normalization layer, ReLU activation layer, and dropout layer. The number of neurons in each layer is 100. Other configurations are the same as NN.                                                                                                                                                                                                                                                                                                           \\
 \midrule
	DNN2 (5-hidden+residual)               & In Deep Neural Networks, a residual connection\cite{he2016deep} creates a shortcut by directly linking two layers, thereby providing a shorter path between them.                                                                                                                                                                                                                                                                                                                                                                                                                                                                                                                                                                                                              & In DNN2, we employ a residual connection to connect the output of the first and fourth layers. Other configurations are the same as DNN1.                                                                                                                                                                                                                                                                                                                                                                                          \\
 \midrule
	DNN3 (9-hidden+residual)               & The DNN3 is configured with 9 hidden layers                                                                                                                                                                                                                                                                                                                                                                                                                                                                                                                                                                                                                                                                                                                        & The DNN3 model is configured with 9 hidden layers with residual connections that link the output of the first hidden layer to the fourth and the output of the fifth hidden layer to the eighth. Other configurations are the same as DNN2.                                                                                                                                                                                                                                                                                        \\
 \bottomrule
\end{longtable}

\subsection{T-test}
\begin{table}[!h]
	\centering
	\begin{tabular}{lcccc}
         \toprule
		\textbf{Model} & \textbf{P Value for Accuracy} & \textbf{P Value for Recall} & \textbf{P Value for Precision} & \textbf{P Value for F1 Score} \\
		\midrule
		LR    & 0.022 & 0.024 & 0.021 & 0.032 \\
		SVM   & 0.034 & 0.041 & 0.036 & 0.034 \\
		DT    & \(1.50 \times 10^{-6}\) & \(1.50 \times 10^{-6}\) & \(2.15 \times 10^{-6}\) & \(1.36 \times 10^{-6}\) \\
		RF    & \(4.98 \times 10^{-6}\) & \(7.70 \times 10^{-6}\) & \(6.42 \times 10^{-6}\) & \(5.84 \times 10^{-6}\) \\
		NN    & 0.003 & 0.007 & 0.008 & 0.004 \\
		DNN1  & 0.021 & 0.020 & 0.024 & 0.015 \\
		DNN2  & 0.042 & 0.078 & 0.014 & 0.087 \\
		DNN3  & 0.023 & 0.157 & 0.042 & 0.175 \\
          \bottomrule
	\end{tabular}
	\caption{Binary Classification Comparison of P Values for Different Models}
	\label{tab:p-value-2class}
\end{table}

\begin{table}[!h]
\centering
\begin{tabular}{lcccc}
\toprule
\textbf{Model} & \textbf{P Value for Accuracy} & \textbf{P Value for Recall} & \textbf{P Value for Precision} & \textbf{P Value for F1 Score} \\
\midrule
LR & 0.048 & 0.045 & 0.041 & 0.070 \\

SVM & 0.018 & 0.026 & 0.031 & 0.011 \\

DT & 0.003 & 0.002 & 0.002 & 0.003 \\

RF & 0.004 & 0.003 & 0.004 & 0.003 \\

NN & \(6.28e \times 10^{-8}\) & \(2.68e \times 10^{-9}\) & \(1.31e \times 10^{-8}\) & \(4.97e \times 10^{-8}\)\\

DNN1 & 0.006 & \(8.27e \times 10^{48}\) & 0.002 & 0.003 \\

DNN2 & 0.043 & 0.048 & 0.010 & 0.053 \\

DNN3 & 0.004 & 0.011 & 0.005 & 0.009 \\
\bottomrule
\end{tabular}
\caption{Ternary Classification Comparison of P Values for Different Models}
\label{tab:p-value-3class}
\end{table}

\subsection{Feature Importance}
\begin{table}[]
	\centering
	\begin{tabular*}{\textwidth}{@{\extracolsep{\fill}}|lr|lr|lr|lr|}
         \toprule
		\multicolumn{2}{c}{2 Class }    & \multicolumn{2}{c}{2 Class Syn}     & \multicolumn{2}{c}{3 Class}     & \multicolumn{2}{c}{3 Class Syn}     \\
          \midrule
		Top-20 features                         & COEF           & Top-20 features                         & COEF             & Top-20 features                        & COEF           & Top-20 features                         & COEF             \\
          \midrule
		P7 NFOD                                 & -1.78          & P7 NFOD                                 & -1.73            & P7 SD                                  & 1.82           & P7 SD                                   & 1.76             \\
		P7 Mobility                             & 1.27           & F7 PSD $\alpha$                         & 1.25             & AF4 Mobility                           & -1.28          & \textbf{AF4 DTW $\alpha$}               & -1.48            \\
		F7 PSD $\alpha$                         & 1.25           & \textbf{AF4 DTW $\alpha$}               & 1.13             & P7 Mobility                            & -1.21          & AF4 Mobility                            & -1.14            \\
		F3 NFOD                                 & -1.22          & F3 NFOD                                 & -1.09            & P7 NFOD                                & 1.09           & F4 DE $\delta$                          & 1.10             \\
		AF3 Mobility                            & 1.06           & P7 Mobility                             & 1.03             & F4 DE $\delta$                         & 1.08           & P7 Mobility                             & -1.05            \\
		F3 Mobility                             & 1.02           & AF3 Mobility                            & 1.03             & F8 SD                                  & 1.01           & \textbf{AF3 DTW $\theta$}               & 0.98             \\
		T7 PSD $\theta$                         & 0.89           & \textbf{AF3 DTW $\theta$}               & -0.99            & F7 NFOD                                & 0.95           & F7 NFOD                                 & 0.98             \\
		T7 Mobility                             & 0.79           & \textbf{F8 DTW $\theta$}                & -0.94            & P7 Variance                            & -0.94          & F8 SD                                   & 0.93             \\
		P7 Variance                             & 0.78           & T7 PSD $\theta$                         & 0.93             & P7 Activity                            & -0.94          & P7 Variance                             & -0.91            \\
		P7 Activity                             & 0.78           & \textbf{F8 DTW $\alpha$}                & -0.91            & F4 Mobility                            & 0.87           & P7 Activity                             & -0.91            \\
		AF3 SD                                  & -0.77          & \textbf{F4 DTW $\alpha$}                & -0.88            & F8 DE FB                               & -0.87          & T7 DE FB                                & -0.90            \\
		AF4 Mobility                            & 0.76           & \textbf{F3 DTW $\beta$}                 & -0.88            & T7 PPM                                 & 0.85           & AF4 DE FB                               & -0.90            \\
		T7 SD                                   & -0.75          & \textbf{AF3 DTW $\alpha$}               & 0.88             & F7 Mobility                            & -0.85          & P7 NFOD                                 & 0.87             \\
		F7 NFOD                                 & -0.74          & T7 SD                                   & -0.83            & F7 SD                                  & 0.84           & F8 DE FB                                & -0.86            \\
		P7 DE FB                                & -0.70          & T7 Mobility                             & 0.82             & AF4 DE FB                              & -0.75          & AF4 PPM                                 & 0.85             \\
		T7 PSD $\alpha$                         & -0.66          & \textbf{AF3 DTW $\delta$}               & 0.82             & F8 PSD $\beta$                         & 0.75           & \textbf{T7 DTW $\delta$}                & -0.83            \\
		F8 PSD $\beta$                          & -0.60          & \textbf{F7 DTW $\delta$}                & -0.81            & F4 SD                                  & -0.74          & F7 Mobility                             & -0.81            \\
		F4 AAFOD                                & 0.60           & \textbf{F4 DTW $\beta$}                 & -0.80            & P7 DE $\delta$                         & -0.74          & F8 PSD $\beta$                          & 0.79             \\
		AF3 NFOD                                & -0.56          & F3 Mobility                             & 0.79             & T7 DE FB                               & -0.72          & \textbf{F4 DTW $\alpha$}                & 0.79             \\
		F8 average                              & 0.56           & T7 PSD $\alpha$                         & -0.77            & AF4 PPM                                & 0.72           & AF4 DE $\delta$                         & -0.76            \\
          \bottomrule
	\end{tabular*}
	\caption{The top-20 important features  selected by sorting the absolute significant coefficient (COEF) values of LR. Features in bold represent inter-brain synchrony features, and Syn represents the model trained on dataset including inter-brain synchrony features. The full names of the features are detailed in section \labelcref{featurename}.}
	\label{Feature LR}
\end{table}

\begin{table}[]
	\centering
	\begin{tabular*}{\textwidth}{@{\extracolsep{\fill}}|lr|lr|lr|lr|}
             \toprule
		\multicolumn{2}{c}{2 Class }    & \multicolumn{2}{c}{2 Class Syn}     & \multicolumn{2}{c}{3 Class}     & \multicolumn{2}{c}{3 Class Syn}     \\
              \midrule
		Top-20 features                  & IMP(\%)        & Top-20 features                                  & IMP(\%)          & Top-20 features                          & IMP(\%)        & Top-20 features                         & IMP(\%)          \\
              \midrule
		F4 LBP $\delta$                  & 1.30           & \textbf{P7 DTW $\alpha$}                         & 3.13             & F4 PSD $\theta$                          & 1.10           & F4 PSD $\theta$                         & 1.38             \\
		F4 Mean PSD                      & 1.20           & \textbf{AF4 CCC $\alpha$}                        & 2.01             & P7 DE $\theta$                           & 1.10           & P7 \textbf{DTW $\alpha$}                & 1.17             \\
		AF4 Kurtosis                     & 1.15           & \textbf{F7 CCC $\beta$}                          & 1.90             & F4 Energy                                & 1.03           & F4 average                              & 1.12             \\
		P7 DE $\alpha$                   & 1.13           & F4 Mean PSD                                      & 1.17             & F4 Power                                 & 0.95           & AF3 Kurtosis                            & 1.01             \\
		F8 Kurtosis                      & 1.09           & \textbf{F3 CCC $\alpha$}                         & 1.07             & F4 average                               & 0.94           & AF3 PSD $\beta$                         & 1.01             \\
		P7 LBP $\theta$                  & 1.09           & \textbf{F3 DTW $\theta$}                         & 1.06             & F4 LBP $\delta$                          & 0.90           & P7 DE $\theta$                          & 0.94             \\
		P7 DE $\theta$                   & 1.07           & \textbf{F7 DTW $\delta$}                         & 1.02             & F4 Mean PSD                              & 0.90           & F8 PPM                                  & 0.92             \\
		F7 PSD $\alpha$                  & 1.03           & \textbf{F3 DTW $\alpha$}                         & 1.01             & AF3 Kurtosis                             & 0.90           & F4 LBP $\delta$                         & 0.88             \\
		P7 LBP $\alpha$                  & 1.01           & \textbf{AF3 CCC $\alpha$}                        & 1.01             & F8 PPM                                   & 0.87           & P7 LBP $\theta$                         & 0.79             \\
		F4 average                       & 0.99           & \textbf{P7 CCC $\alpha$}                         & 0.99             & AF3 PSD $\beta$                          & 0.83           & F4 Energy                               & 0.78             \\
		T7 PSD $\theta$                  & 0.95           & \textbf{AF4 CCC $\beta$}                         & 0.98             & P7 LBP $\theta$                          & 0.80           & F8 Kurtosis                             & 0.76             \\
		T7 Kurtosis                      & 0.93           & \textbf{F8 CCC $\alpha$}                         & 0.97             & F4 LBP $\alpha$                          & 0.80           & F4 Power                                & 0.75             \\
		F4 Energy                        & 0.92           & \textbf{F8 CCC $\beta$}                          & 0.90             & P7 Kurtosis                              & 0.76           & \textbf{P7 CCC $\beta$}                 & 0.70             \\
		F8 PSD $\beta$                   & 0.90           & \textbf{F8 CCC $\delta$}                         & 0.87             & AF3 LBP $\beta$                          & 0.75           & \textbf{AF4 CCC $\alpha$}               & 0.66             \\
		F4 Power                         & 0.90           & \textbf{F4 CCC $\alpha$}                         & 0.87             & AF3 AAFOD                                & 0.75           & \textbf{F8 CCC $\alpha$}                & 0.65             \\
		P7 Kurtosis                      & 0.90           & \textbf{AF4 CCC $\theta$}                        & 0.84             & F4 DE $\alpha$                           & 0.75           & F7 PSD $\beta$                          & 0.65             \\
		F4 PSD $\theta$                  & 0.88           & \textbf{F7 CCC $\delta$}                         & 0.81             & F7 LBP $\alpha$                          & 0.74           & F4 LBP $\alpha$                         & 0.64             \\
		F4 PSD $\delta$                  & 0.85           & F4 average                                       & 0.79             & AF3 DE $\beta$                           & 0.72           & \textbf{T7 DTW $\theta$}                & 0.62             \\
		T7 HOZC                          & 0.80           & F4 PSD $\delta$                                  & 0.78             & AF4 DE $\beta$                           & 0.69           & F4 DE $\alpha$                          & 0.62             \\
		AF4 PSD $\alpha$                 & 0.80           & \textbf{P7 CCC $\theta$}                         & 0.77             & F8 Kurtosis                              & 0.69           & \textbf{T7 DTW $\beta$}                 & 0.61             \\
              \bottomrule
	\end{tabular*}
	\caption{The top-20 important features selected by sorting the absolute significant importance (IMP) values of RF. Features in bold represent inter-brain synchrony features, and Syn represents the model trained on dataset including inter-brain synchrony features. The full names of the features are detailed in section \labelcref{featurename}.}
	\label{Feature RF}
\end{table}

\begin{table}[]
	\centering
	\begin{tabular*}{\textwidth}{@{\extracolsep{\fill}}|lr|lr|lr|lr|}
             \toprule
		\multicolumn{2}{c}{2 Class }    & \multicolumn{2}{c}{2 Class Syn}     & \multicolumn{2}{c}{3 Class}     & \multicolumn{2}{c}{3 Class Syn}     \\
              \midrule
		Top-20 features                  & SHAP       & Top-20 features                                  & SHAP          & Top-20 features                          & SHAP        & Top-20 features                         & SHAP         \\
              \midrule
                F7 PSD $\alpha$     & 2.89  & F7 PSD $\alpha$               & 5.34  & AF3 AAFOD         & 4.88  & P7 PSD $\beta$            & -7.72  \\ 
                AF3 Mobility        & -2.55 & \textbf{F8 CCC $\delta$}      & -4.72 & F7 PSD $\alpha$   & -4.26 & F7 PSD $\beta$            & -4.68  \\ 
                T7 Activity         & -2.46 & F7 PSD $\theta$               & 4.63  & AF3 PSD $\beta$   & 3.94  & \textbf{F8 DTW $\delta$}  & 4.63  \\ 
                F7 mean PSD         & -2.23 & \textbf{AF3 DTW $\delta$}     & -4.14 & AF3 DE $\beta$    & -3.64 & P7 Kurtosis               & -4.53  \\ 
                F8 Activity         & 2.20  & \textbf{F8 CCC $\alpha$}      & -3.50 & F4 AAFOD          & -3.11 & F8 PSD $\theta$           & 4.09  \\ 
                P7 DE FB            & -2.00 & F4 LBP $\delta$               & 3.44  & F8 AAFOD          & -3.11 & AF4 PSD $\delta$          & -4.08  \\ 
                F4 DE $\alpha$      & -1.67 & F4 AAFOD                      & -3.38 & F3 PSD $\alpha$   & 3.06  & AF3 LBP $\beta$           & -3.97  \\ 
                F7 PSD $\theta$     & 1.62  & \textbf{F3 DTW $\delta$}      & 3.25  & P7 AAFOD          & 2.92  & AF3 PSD $\beta$           & -3.87  \\ 
                T7 Kurtosis         & -1.62 & P7 Variance                   & -3.15 & AF3 Kurtosis      & 2.91  & AF4 AAFOD                 & 3.86  \\ 
                T7 Mobility         & -1.52 & \textbf{F4 DTW $\theta$}      & -3.08 & F7 mean PSD       & -2.86 & F4 DE $\delta$            & -3.81  \\ 
                P7 AAFOD            & 1.47  & \textbf{F3 CCC $\theta$}      & -2.84 & F7 PSD $\beta$    & -2.73 & P7 HOZC                   & -3.77  \\ 
                F3 HOZC             & -1.33 & AF3 PPM                       & 2.74  & T7 DE $\delta$    & 2.61  & \textbf{F4 DTW $\delta$}  & -3.76  \\ 
                F8 PSD $\alpha$     & -1.19 & \textbf{P7 DTW $\beta$}       & 2.70  & AF4 LBP $\alpha$  & -2.47 & F3 LBP $\theta$           & 3.69  \\ 
                T7 PSD $\beta$      & 1.17  & T7 PSD $\alpha$               & -2.65 & AF3 PSD $\alpha$  & -2.41 & T7 PSD $\theta$           & 3.63  \\ 
                F8 Variance         & 1.11  & F4 LBP $\theta$               & -2.64 & F4 LBP $\alpha$   & -2.15 & T7 DE FB                  & -3.50 \\ 
                AF4 Variance        & 1.10  & P7 PSD $\beta$                & -2.61 & F8 DE $\beta$     & 2.03  & F3 DE $\theta$            & 3.39  \\ 
                F3 NFOD             & -1.07 & \textbf{F8 DTW $\theta$}      & 2.57  & F4 Energy         & 1.97  & T7 PSD $\alpha$           & 3.26  \\ 
                F7 PSD $\beta$      & 1.01  & F8 PPM                        & 2.51  & F3 DE $\beta$     & 1.95  & \textbf{AF3 DTW $\alpha$} & 3.20 \\ 
                F8 PSD $\beta$      & 1.01  & F4 Energy                     & -2.48 & F7 PSD $\theta$   & 1.93  & P7 PSD $\alpha$           & 3.12  \\ 
                AF4 Mobility        & 1.00  & P7 PPM                        & -2.40 & AF4 PPM           & -1.89 & \textbf{F7 CCC $\delta$}  & -3.06  \\ 
              \bottomrule
	\end{tabular*}
	\caption{The top-20 important features selected by sorting the absolute SHapley Additive exPlanations (SHAP) values of NN. Features in bold represent inter-brain synchrony features, and Syn represents the model trained on dataset including inter-brain synchrony features. The full names of the features are detailed in section \labelcref{featurename}. In binary classification tasks, the unit of SHAP value is $10^{-9}$, and in ternary classification tasks, the unit of SHAP value is $10^{-8}$.}
	\label{Feature NN}
\end{table}

\begin{table}[]
	\centering
	\begin{tabular*}{\textwidth}{@{\extracolsep{\fill}}|lr|lr|lr|lr|}
             \toprule
		\multicolumn{2}{c}{2 Class }    & \multicolumn{2}{c}{2 Class Syn}     & \multicolumn{2}{c}{3 Class}     & \multicolumn{2}{c}{3 Class Syn}     \\
              \midrule
		Top-20 features                  & SHAP        & Top-20 features                                  & SHAP          & Top-20 features                          & SHAP        & Top-20 features                         & SHAP          \\
              \midrule
                        P7 DE FB  & 2.21  & \textbf{F7 CCC $\delta$}      & 1.14  & T7 Kurtosis       & -2.13 & \textbf{AF4 CCC $\theta$}     & -2.27  \\ 
                        F4 PSD $\delta$ & 1.92  & AF4 HOZC                      & -1.12 & AF4 LBP $\alpha$  & -1.90 & P7 Kurtosis                   & -2.22  \\ 
                        P7 DE $\beta$   & -1.82 & \textbf{F3 CCC $\theta$}      & 1.08  & T7 HOZC           & 1.81  & AF3 Kurtosis                  & -2.17  \\ 
                        F3 LBP $\theta$ & -1.81 & T7 DE $\delta$                & 1.06  & AF4 DE $\alpha$   & -1.74 & T7 LBP $\beta$                & -1.63  \\ 
                        F4 LBP $\beta$  & 1.50  & \textbf{AF4 CCC $\beta$}      & 0.98  & F7 AAFOD          & -1.57 & F4 \textbf{CCC $\theta$}      & -1.43  \\ 
                        F3 Activity     & 1.48  & T7 PSD $\beta$                & -0.98 & T7 Mobility       & -1.51 & F7 DE $\beta$                 & 1.41  \\ 
                        F7 Kurtosis     & 1.43  & F4 LBP $\beta$                & 0.95  & F3 LBP $\beta$    & 1.50  & F7 DE FB                      & -1.39  \\ 
                        F3 Variance     & 1.41  & \textbf{F8 CCC $\delta$}      & 0.93  & F7 DE $\alpha$    & -1.40 & F7 Kurtosis                   & -1.38  \\ 
                        F4 mean PSD     & 1.37  & \textbf{F7 CCC $\alpha$}      & -0.91 & F7 LBP $\alpha$   & -1.37 & F4 LBP $\beta$                & -1.30 \\ 
                        P7 PPM          & -1.35 & AF4 Energy                    & -0.90 & F7 HOZC           & -1.35 & AF3 DE $\theta$               & 1.26  \\ 
                        F4 LBP $\alpha$ & -1.29 & F4 HOZC                       & -0.88 & T7 DE $\theta$    & 1.27  & F3 NOFD                       & -1.25  \\ 
                        F3 Energy       & 1.23  & T7 AAFOD                      & 0.87  & T7 DE $\beta$     & 1.23  & \textbf{T7 CCC $\theta$}      & -1.24  \\ 
                        F8 DE $\alpha$  & -1.21 & AF3 DE FB                     & -0.85 & F3 PSD $\delta$   & -1.18 & P7 DE FB                      & 1.19  \\ 
                        AF4 Energy      & -1.20 & F3 DE $\theta$                & -0.74 & F8 HOZC           & 1.18  & F4 LBP $\theta$               & -1.16  \\ 
                        F3 PSD $\delta$ & 1.15  & \textbf{AF3 CCC $\theta$}     & -0.74 & F4 mean PSD       & -1.16 & \textbf{T7 CCC $\delta$}      & 1.12  \\ 
                        AF4 DE $\delta$ & 1.15  & AF3 PSD $\delta$              & -0.74 & P7 Kurtosis       & 1.14  & \textbf{F3 CCC $\alpha$}      & -1.11  \\ 
                        F3 DE FB        & -1.14 & F7 DE $\alpha$                & -0.72 & AF3 HOZC          & -1.12 & \textbf{F4 DTW $\delta$}      & -1.10 \\ 
                        F3 Mobility     & 1.11  & T7 Mobility                   & -0.69 & P7 LBP $\theta$   & -1.08 & \textbf{F7 CCC $\delta$}      & -1.08  \\ 
                        AF3 SD          & -1.10 & \textbf{AF3 CCC $\beta$}      & 0.67  & F3 DE $\alpha$    & -1.07 & T7 LBP $\theta$               & 1.07  \\ 
                        F7 PPM          & -1.09 & F8 Mobility                   & -0.65 & F3 PSD $\beta$    & -1.06 & T7 AAFOD                      & -1.06  \\ 
              \bottomrule
	\end{tabular*}
	\caption{The top-20 important features selected by sorting the absolute SHapley Additive exPlanations (SHAP)  values of DNN3. Features in bold represent inter-brain synchrony features, and Syn represents the model trained on dataset including inter-brain synchrony features. The full names of the features are detailed in section \labelcref{featurename}. The unit of SHAP value is $10^{-8}$ both in binary and ternary classification tasks.}
	\label{Feature DNN3}
\end{table}

\end{document}